\documentclass[aps,superscriptaddress,nofootinbib,floatfix,epfs,preprintnumbers]{revtex4}
\usepackage{amsfonts,amscd,amsmath,amssymb,graphicx,color,float}
\usepackage[section]{placeins}
\def\ep{\text{e}}
\def\g{\mathsf{g}}
\def\oh{\frac{1}{2}}
\def\s{\mathsf{s}}

\def\k{\mathsf{k}}
\def\n{\mathsf{n}}
\def\km{-\frac{1}{4}\ep^{\frac{1}{4}}}
\def\vz{v_{\text{\tiny 0}}}
\def\vo{v_{\text{\tiny 1}}}
\def\Vz{{\text v}_{\text{\tiny 0}}}
\def\Vo{{\text v}_{\text{\tiny 1}}}

\def\rq{r_q}
\def\rqb{r_{\bar q}}
\def\rv{r_v}
\def\rvb{r_{\bar v}}
\def\rqb{r_{\bar q}}

\def\2Qqb{\text{\tiny QQ}\bar{\text{\tiny q}}\bar{\text{\tiny q}}}
\def\Qqqb{\bar{\text{\tiny Q}}\bar{\text{\tiny q}}\bar{\text{\tiny q}}}
\def\Qqqq{\text{\tiny Qq}\bar{\text{\tiny q}}\bar{\text{\tiny q}}}
\def\Qqb{\text{\tiny Q}\bar{\text{\tiny q}}}

\def\QQb{\text{\tiny Q}\bar{\text{\tiny Q}}}
\def\Qqq{\text{\tiny Qqq}}
\def\Qqqb{\bar{\text{\tiny Q}}\bar{\text{\tiny q}}\bar{\text{\tiny q}}}
\def\QQq{\text{\tiny QQq}}
\def\QQ{\text{\tiny QQ}}
\def\3Q{3\text{\tiny Q}}
\def\qQb{\text{\tiny q}\bar{\text{\tiny Q}}}
\def\qqq{3\bar{\text{\tiny q}}}

\def\EI{E^{\text{\tiny (I)}}_{\QQq}}

\begin{document}
\preprint{LMU-ASC 40/21}
\title{On the $QQ\bar q\bar q$-Quark Potential in String Models}
\author{Oleg Andreev}
 \affiliation{L.D. Landau Institute for Theoretical Physics, Kosygina 2, 119334 Moscow, Russia}
\affiliation{Arnold Sommerfeld Center for Theoretical Physics, LMU-M\"unchen, Theresienstrasse 37, 80333 M\"unchen, Germany}
\begin{abstract} 
We propose a string theory construction for the system of two heavy quarks and two light antiquarks. The potential of the system is a function of separation between the quarks. We define a critical separation distance below which the system can be thought of mainly as a compact tetraquark. The results show the universality of the string tension and factorization at small separations expected from heavy quark-diquark symmetry. Our estimate of the screening length is in the range of lattice QCD. We also make a comparison with the potential of the $QQq$ system. The potentials look very similar at small quark separations but at larger separations they differ. The reason for this is that the flattening of the potentials happens at two well-separated scales as follows from the two different mechanisms: string breaking by light quarks for $QQq$ and string junction annihilation for $QQ\bar q\bar q$. Moreover, a similar construction can also be applied to the $\bar Q\bar Q qq$ system.
\end{abstract}
\maketitle
\section{Introduction}
\renewcommand{\theequation}{1.\arabic{equation}}
\setcounter{equation}{0}

Since the proposal of the quark model by Gell-Mann \cite{GM} and Zweig \cite{zweig} in the sixties, exotic hadrons remain a challenge for the physics of strong interactions \cite{ali}. Recently the LHCb Collaboration has announced the discovery of a double charm exotic meson $T_{cc}^+ = cc\bar u\bar d$ with a mass around $3875\,\text{MeV}$ \cite{LHCb}. This revived and reinforced the old interest \cite{JMR0} in the search for a theoretical description of such doubly heavy hadrons (tetraquarks).  

Although lattice gauge theory is one of the basic tools for studying nonperturbative phenomena in QCD, with increasing progress in the study of the doubly heavy tetraquarks \cite{AF}, the need to understand the physics behind computational complexity forces one to employ string models. A special class of these called holographic (AdS/QCD) models has received much attention in the last years. The hope is that the gauge/string duality does provide new theoretical tools for studying strongly couple gauge theories.\footnote{For the further development of these ideas in the context of QCD, see the book \cite{book-u} and references therein.} In those models the string configurations for tetraquarks were qualitatively discussed in \cite{a-3q0, sw}. Making it more precise for the case of doubly heavy tetraquarks will be one of the goals of the present paper. 

In this paper, we propose a string theory construction for the $QQ\bar q\bar q$ system which allows us to compute its minimal energy (potential). So far there have been no such constructions in the literature. We follow the standard hadro-quarkonium picture \cite{voloshin} so that the heavy quark pair is considered as being embedded in light antiquark clouds, and assume that the heavy quarks are heavy enough to be well approximated as static. The potential is determined by the relative separation between the quarks. Then one can use it as an input to the potential models to find the bound states of the system. 

In Sec.II, we briefly review the framework in which we will work, and recall some previous results. In Sec.III, we construct and analyze a set of string configurations describing the $QQ\bar q\bar q$ system. Then among those we find the configurations which contribute to the potential of the system. This enables us to reconstruct the potential along the lines of lattice QCD. We also introduce a critical separation distance $\ell_{\2Qqb}$ between the heavy quarks which is associated with the transition between the dominate configurations. For separations less than $\ell_{\2Qqb}$ the connected configuration is dominant whilst for greater than $\ell_{\2Qqb}$ the disconnected one. We then go on in Sec. IV to discuss the relation between the potentials $V_{\QQq}$ and $V_{\2Qqb}$ and to compare our results with those on the lattice. We conclude in Sec.V by making a few comments. Appendix A presents our notation and definitions. To make the paper more self-contained, we include the necessary results and technical details in Appendices B and C.

\section{Preliminaries}
\renewcommand{\theequation}{2.\arabic{equation}}
\setcounter{equation}{0}

In our discussion we will use the formalism developed in \cite{a-strb1}. We illustrate most ideas with one of the simplest AdS/QCD models which purports to mimic QCD with two light flavors, but the extension to other models is straightforward. 

First let us specify a five dimensional geometry. The metric is taken to be of the form 

\begin{equation}\label{metric}
ds^2=\ep^{\s r^2}\frac{R^2}{r^2}\Bigl(dt^2+d\vec x^2+dr^2\Bigr)
\,.
\end{equation}
Such geometry is a deformation of the Euclidean $\text{AdS}_5$ space of radius $R$, with a deformation parameter $\s$. So, it has a boundary at $r=0$. Two features make it especially attractive: computational simplicity and phenomenological applications.\footnote{For some of those, see \cite{az1,a-hyb,a-3qPRD}.} In addition, we introduce a background scalar field $\text{T}(r)$ which describes light (anti) quarks at string endpoints in the interior of five-dimensional space \cite{son}. This enables one to construct disconnected string configurations, and in particular to model the phenomenon of string breaking. We introduce a single scalar field (tachyon), since in what follows we consider only the case of two light quarks of equal mass.\footnote{The use of the term tachyon seems particularly appropriate in virtue of instability of a QCD string and the worldsheet coupling to the tachyon (see \eqref{Sq}).}

Just as for Feynman diagrams in field theory, we need the building blocks to construct string configurations. The first one is a Nambu-Goto string whose action is 

\begin{equation}\label{NG}
S_{\text{\tiny NG}}=\frac{1}{2\pi\alpha'}\int d^2\xi\,\sqrt{\gamma^{(2)}}
\,.
\end{equation}
Here $\gamma$ is an induced metric, $\alpha'$ is a string parameter, and $\xi^i$ are world-sheet coordinates. 

The second is a pair of string junctions, called in nowadays the baryon vertices, at which three strings meet. In the AdS/CFT correspondence the baryon vertex is supposed to be a five brane wrapped on an internal space $\mathbf{X}$ and correspondingly the antibaryon vertex an antibrane \cite{witten}. Those both look point-like in five dimensions. In \cite{a-3q} it was observed that the action for the baryon vertex, written in the static gauge, 

\begin{equation}\label{baryon-v}
S_{\text{v}}=\tau_v\int dt \,\frac{\ep^{-2\s r^2}}{r}
\,
\end{equation}
yields very satisfactory results, when compared to the lattice calculations of the three quark potential. In fact, this action is given by the volume of the brane if $\tau_v={\cal T}_5R\,\text{vol}(\mathbf{X})$, with ${\cal T}_5$ a brane tension. Unlike AdS/CFT, we treat $\tau_v$ as a free parameter to somehow account for $\alpha'$-corrections as well as possible impact of the other background fields. It is natural to also take the action \eqref{baryon-v} for the antibaryon vertex so that $S_{\bar{\text{v}}}=S_{\text{v}}$.

 The third building block, which takes account of light quarks at string endpoints, is provided by the scalar field. It couples to the worldsheet boundary as an open string tachyon $S_{\text{q}}=\int d\tau e\,\text{T}$, where $\tau$ is a coordinate on the boundary and $e$ is a boundary metric. In what follows, we consider only a constant field $\text{T}_0$ and worldsheets whose boundaries are lines in the $t$ direction. In that case, the action written in the static gauge is    

\begin{equation}\label{Sq}
S_{\text q}=\text{T}_0R\int dt \frac{\ep^{\frac{\s}{2}r^2}}{r}
\,.
\end{equation}
It is nothing else than the action of a point particle of mass ${\text T}_0$ at rest. Clearly, the same action also describes the light antiquarks at string endpoints, and hence $S_{\bar{\text q}}=S_{\text q}$. 

In fact, there is a visual analogy between tree level Feynman diagrams and static string configurations. In the language of Feynman diagrams the above building blocks play respectively the roles of propagators, vertices and tadpoles.

\section{The $QQ\bar q\bar q$-Quark Potential via Gauge/String Duality}
\renewcommand{\theequation}{3.\arabic{equation}}
\setcounter{equation}{0}

Now we will begin our discussion of the $QQ\bar q\bar q$ system. In doing so, we follow the hadro-quarkonium picture \cite{voloshin} and hence think of the light antiquarks as clouds.\footnote{Clearly, it does not make a lot of sense to speak about the positions of the antiquarks. One can only do so in terms of their average positions or, equivalently, the centers of the clouds. We keep that in mind every time we speak about the light antiquarks.} The heavy quarks are point-like objects inside the clouds. Our goal is to determine the potential as a function of separation between the quarks. We start our discussion with a connected string configuration, then continue with disconnected ones, and finally end up with the potential.

\subsection{A connected string configuration}

An intuitive way to see the right configuration in five dimensions is to place the standard tetraquark configuration (two quarks and antiquarks connected by strings, as usual in four dimensions) on the boundary of five-dimensional space. A gravitational force pulls the light antiquarks and strings into the interior, whereas the heavy (static) quarks remain at rest. As a result, the configuration we are looking for takes the form of one of those shown in Figures \ref{QQqq1}-\ref{QQqq4}. 

It is natural to suggest that the string configuration for the ground state is dictated by symmetry. If so, then there are the two most symmetric cases: 1. The antiquarks are in the middle between the quarks. 2. The each antiquark sits on top of one of the quarks. The configurations shown in Figure \ref{QQqq1}-\ref{QQqq4} correspond to the first case and, as we will see shortly, one of the disconnected configurations of Figure \ref{disc} to the second.

\subsubsection{Small $\ell$}

\begin{figure}[htbp]
\centering
\includegraphics[width=6.25cm]{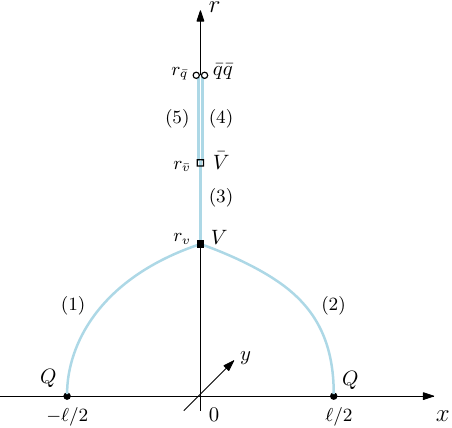}
\caption{{\small A static string configuration at small heavy quark separations. The heavy quarks $Q$ are placed on the boundary, while the light antiquarks $\bar q$, baryon $V$ and antibaryon $\bar V$ vertices in the bulk of the five-dimensional space.}}
\label{QQqq1}
\end{figure}

For this case, the corresponding string configuration is presented in Figure \ref{QQqq1}.\footnote{This follows from the analysis of subsection 5.} The total 
action is the sum of the Nambu-Goto actions plus the actions for the vertices and antiquarks

\begin{equation}\label{S1}
S=\sum_{i=1}^5 S_{\text{\tiny NG}}^{(i)}+2S_{\text{v}}+2S_{\text q}
\,.
\end{equation}
If one picks the static gauge $\xi^1=t$ and $\xi^2=r$ for the Nambu-Goto actions, then the boundary conditions for the $x$'s are  

\begin{equation}\label{boundaryc}
x^{(1)}(0)=-\oh\ell\,,\qquad
x^{(2)}(0)=\oh\ell\,,\qquad
x^{(1,2,3)}(\rv)=x^{(3,4,5)}(\rvb)=x^{(4,5)}(\rqb)=0\,,
\end{equation}
and the action becomes\footnote{We drop the subscript $(i)$ when it does not cause confusion.} 

\begin{equation}\label{Sconf-s}
S=\g T
\biggl(2\int_{0}^{\rv} \frac{dr}{r^2}\,\ep^{\s r^2}\sqrt{1+(\partial_r x)^2}\,\,+\int_{\rv}^{\rvb} \frac{dr}{r^2}\,\ep^{\s r^2}\,\,+2\int_{\rvb}^{\rqb} \frac{dr}{r^2}\,\ep^{\s r^2}
+3\k\,\frac{\ep^{-2\s\rv^2}}{\rv}
+3\k\,\frac{\ep^{-2\s\rvb^2}}{\rvb}
+2\n\frac{\ep^{\frac{1}{2}\s\rqb^2}}{\rqb}
\,\biggr)
\,.
\end{equation}
Here $\k=\frac{\tau_v}{3\g}$, $\n=\frac{\text{T}_0R}{\g}$, $\partial_rx=\frac{\partial x}{\partial r}$, and $T=\int_0^T dt$. We set $x=0$ for all the straight strings. 

Using the formulas from Appendix B for the case $\alpha\geq 0$, we immediately deduce that

\begin{equation}\label{lsmall}
\ell=\frac{2}{\sqrt{\s}}{\cal L}^+(\alpha,v)
\,
\end{equation}
and the energy of the configuration is 

\begin{equation}\label{Esmall}
E_{\2Qqb}=\g\sqrt{\s}
\biggl(2{\cal E}^+(\alpha,v)
+
2{\cal Q}(\bar q)-{\cal Q}(\bar v)-{\cal Q}(v)
+
3\k\frac{\ep^{-2v}}{\sqrt{v}}
+
3\k\frac{\ep^{-2{\bar v}}}{\sqrt{\bar v}}
+
2\n\frac{\ep^{\oh\bar q}}{\sqrt{\bar q}}
\,\biggr)
+2c
\,.
\end{equation}
Here $v=\s\rv^2$, $\bar v=\s\rvb^2$, $\bar q=\s\rqb^2$, and $\alpha$ is the tangent angle defined in Appendix B.

We still have to extremize the action with respect to the positions of the vertices and light antiquarks. This will provide us with the recipes for gluing the string endpoints together at the vertices and attaching the antiquarks to the string endpoints in the bulk. The physical meaning of those is that the net forces exerted on the vertices and antiquarks must vanish in equilibrium.  

By varying the action with respect to $\rv$, one can deduce that\footnote{In doing so, one has to keep in mind the boundary conditions \eqref{boundaryc}.} 

\begin{equation}\label{alphas1}
2\sin\alpha-1-3\k(1+4v)\ep^{-3v}=0
\,.
\end{equation}
The variation of the action with respect to $\rvb$ results in 

\begin{equation}\label{vb}
1+3\k(1+4{\bar v})\ep^{-3{\bar v}}=0
\,,
\end{equation}
which is the special case of $\eqref{alphas1}$ when $\alpha=0$. A noteworthy fact is that this equation has solutions in the interval $[0,1]$ if and only if $-\frac{\ep^3}{15}\leq\k\leq\km$ \cite{a-QQq}. In particular, for $\k=\km$ the solution is simply $\bar v=\tfrac{1}{12}$. The last equation 

\begin{equation}\label{qb}
\ep^{\frac{\bar q}{2}}+\n(\bar q-1)=0
\,
\end{equation}
comes by varying $\rqb$. It formally coincides with equation \eqref{q} derived in \cite{a-strb1} for light quarks, as should be at zero baryon chemical potential. 

Thus, the energy of the configuration is given in parametric form by $E_{\2Qqb}=E_{\2Qqb}(v)$ and $\ell=\ell(v)$. The parameter goes from $0$ to $\bar v$. Here $\bar v$ is a solution of Eq.\eqref{vb} on the interval $[0,1]$.

\subsubsection{Slightly larger $\ell$}

\begin{figure}[htbp]
\centering
\includegraphics[width=6.5cm]{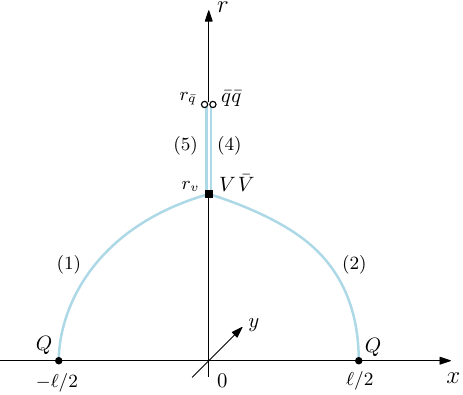}
\caption{{\small A static configuration with the vertices located at the same point on the $r$-axis.}}
\label{QQqq2}
\end{figure}

A simple numerical analysis of \eqref{lsmall} shows that $\ell(v)$ is an increasing function which is finite at $v=\bar v$. This means that as the separation between the heavy quarks is increased, the baryon vertex goes deeper into the bulk until it reaches the antibaryon vertex whose position is independent of the quark separation. As a result, the configuration becomes that of Figure \ref{QQqq2}. It can be thought of as the  
previous one with string (3) shrunk to a point. 

The total action is now given by 

\begin{equation}\label{S2}
S=\sum_{i=1,\,i\not =3}^5 S_{\text{\tiny NG}}^{(i)}+2S_{\text{v}}+2S_{\text q}
\,.
\end{equation}
We choose the same static gauge as before. Then the boundary conditions are 

\begin{equation}\label{boundaryc2}
x^{(1)}(0)=-\oh\ell\,,\qquad
x^{(2)}(0)=\oh\ell\,,\qquad
x^{(i)}(\rv)=x^{(4,5)}(\rqb)=0\,.
\end{equation}
 With these boundary conditions, the action takes the form 

\begin{equation}\label{Sconf-s2}
S=2\g T
\biggl(\int_{0}^{\rv} \frac{dr}{r^2}\,\ep^{\s r^2}\sqrt{1+(\partial_r x)^2}\,\,
+\int_{\rv}^{\rqb} \frac{dr}{r^2}\,\ep^{\s r^2}
+3\k\,\frac{\ep^{-2\s\rv^2}}{\rv}
+\n\frac{\ep^{\frac{1}{2}\s\rq^2}}{\rq}
\,\biggr)
\,.
\end{equation}

Clearly, $\ell$ is given by Eq.\eqref{lsmall} and $E_{\2Qqb}$ by 

\begin{equation}\label{Esmall2}
E_{\2Qqb}=2\g\sqrt{\s}
\biggl(
{\cal E}^+(\alpha,v)
+
{\cal Q}(\bar q)-{\cal Q}(v)
+
\n\frac{\ep^{\oh\bar q}}{\sqrt{\bar q}}
+
3\k\frac{\ep^{-2v}}{\sqrt{v}}
\biggr)
+2c
\,.
\end{equation}
Varying the action \eqref{Sconf-s2} with respect to $\rqb$ leads to Eq.\eqref{qb} and with respect to $\rv$ to 

\begin{equation}\label{alpha2}
\sin\alpha-1-3\k(1+4v)\ep^{-3v}=0
\,.
\end{equation}

So, the energy of the configuration is given parametrically by $E_{\2Qqb}=E_{\2Qqb}(v)$ and $\ell=\ell(v)$ with the parameter $v$ varying from $\bar v$ to $\bar q$. Here $\bar q$ is a solution of Eq.\eqref{qb} on the interval $[0,1]$. 

\subsubsection{Intermediate $\ell$}

\begin{figure}[htbp]
\centering
\includegraphics[width=7.3cm]{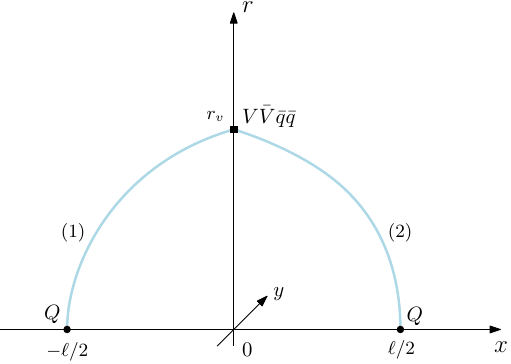}
\caption{{\small A static string configuration at intermediate heavy quark separations. The baryon vertices and antiquarks are at the same point on the $r$-axis. The tangent angle $\alpha$ is non-negative at $r=\rv$.}}
\label{QQqq3}
\end{figure}

Again a numerical analysis shows that $\ell(v)$ is finite at $v=\bar q$ where the vertices reach the light antiquarks. So, to get further, we must consider the configuration of Figure \ref{QQqq3}. One can think of it as two strings meeting at a point-like 

\noindent object (defect made of the vertices and antiquarks) in the bulk.\footnote{It is noteworthy that such a defect results in a cusp formation in the $r$-direction.} So the total action simplifies to 

\begin{equation}\label{S3}
S=\sum_{i=1}^2 S_{\text{\tiny NG}}^{(i)}+2S_{\text{v}}+2S_{\text q}
\,.	
\end{equation}

Choosing the static gauge in the Nambu-Goto actions as before, we consider the $x$'s as a function of $r$ subject to the boundary conditions 

\begin{equation}\label{boundaryc3}
x^{(1)}(0)=-\oh\ell\,,\qquad
x^{(2)}(0)=\oh\ell\,,\qquad
x^{(i)}(\rv)=0\,.
\end{equation}
The total action is then

\begin{equation}\label{Sconf-s3}
S=2\g T
\biggl(\int_{0}^{\rv} \frac{dr}{r^2}\,\ep^{\s r^2}\sqrt{1+(\partial_r x)^2}\,\,
+3\k\,\frac{\ep^{-2\s\rv^2}}{\rv}
+\n\frac{\ep^{\frac{1}{2}\s\rv^2}}{\rv}
\,\biggr)
\,.
\end{equation}

Since the tangent angle at $r=\rv$ is non-negative, the formula \eqref{lsmall} for the distance between the quarks holds. The energy of the configuration is 

\begin{equation}\label{Eint}
E_{\2Qqb}=2\g\sqrt{\s}
\biggl(
{\cal E}^+(\alpha,v)
+
\frac{\n\ep^{\oh v}+3\k\ep^{-2v}}{\sqrt{v}}
\biggr)
+2c
\,.
\end{equation}
It can be simply deduced from \eqref{Esmall2} by taking $\bar q=v$. 

By varying the action with respect to $\rv$, we get 

\begin{equation}\label{alphaint}
\sin\alpha-3\k(1+4v)\ep^{-3v}+\n(v-1)\ep^{-\oh v}=0
\,.
\end{equation}
This is nothing else but the force balance equation at $r=\rv$. For the parameter values we are using, $\alpha$ turns out to be a decreasing function of $v$. It vanishes at $v=\vz$ which is a solution to the equation 

\begin{equation}\label{v0}
3\k(1+4v)\ep^{-3v}+\n(1-v)\ep^{-\oh v}=0
\,.
\end{equation}

In summary, at intermediate quark separations the energy is given by the parametric equations \eqref{lsmall} and \eqref{Eint} with the parameter $v$ varying from $q$ to $\vz$. 

\subsubsection{Large $\ell$ }

\begin{figure}[htbp]
\centering
\includegraphics[width=7.5cm]{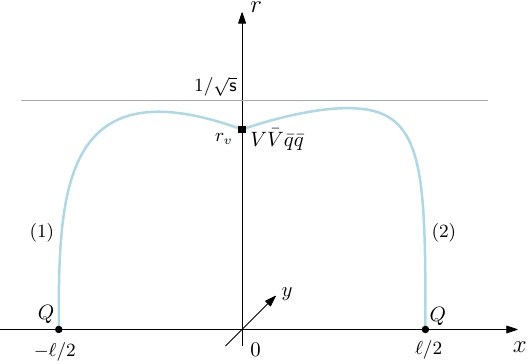}
\caption{{\small A static string configuration at large heavy quark separations. The horizontal line represents the soft wall. The tangent angle at $r=\rv$ is negative.}}
\label{QQqq4}
\end{figure}

From the expression \eqref{lsmall}, it follows that $\ell$ remains finite at $v=\vz$.\footnote{The argument assumes that $\ell(v)$ is an increasing function. This is indeed the case for the model parameter values we are using.} The question arises what is going to happen for larger values of $\ell$? The answer is that $\alpha$ changes the sign from positive to negative so that the configuration profile becomes convex near $x=0$, as shown in Figure \ref{QQqq4}. The strings continue to go deeper in the bulk until finally reach  

\noindent the soft-wall at $r=1/\sqrt{\s}$. The limiting case corresponds to infinite separation between the heavy quarks. It will be described in more detail in subsection 5. 

The configuration is again governed by the total action \eqref{S3}. The expressions for the distance between the quarks and energy are simply obtained by respectively replacing ${\cal L}^+$ and ${\cal E}^+$ with ${\cal L}^-$ and ${\cal E}^-$ as follows from the analysis in Appendix B. So, we have 

\begin{equation}\label{l4}
\ell=
\frac{2}{\sqrt{\s}}{\cal L}^-(\lambda,v)
\,
\end{equation}
and 
\begin{equation}\label{E4}
E_{\2Qqb}= 2\g\sqrt{\s}
\biggl(
{\cal E}^-(\lambda,v)
+
\frac{\n\ep^{\oh v}+3\k\ep^{-2v}}{\sqrt{v}}
\biggr)
+2c
\,.
\end{equation}

The force balance equation at the point $r=\rv$ is given by Eq.\eqref{alphaint}, but now with negative $\alpha$ values. Combining this equation with \eqref{lambda}, we get  

\begin{equation}\label{lambdav}
\lambda(v)=-\text{ProductLog}\biggl[-v\ep^{-v}
\biggl(1-\Bigl(3\k(1+4v)\ep^{-3v}+\n(1-v)\ep^{-\oh v}\Bigr)^2\biggr)^{-\oh}
\biggr]
\,.	
\end{equation}
The limiting value of $v$, when $\lambda$ approaches $1$ and the strings the soft-wall, is determined from the equation

\begin{equation}\label{v1}
\sqrt{1-v^2\ep^{2(1-v)}}+3\k(1+4v)\ep^{-3v}+\n(1-v)\ep^{-\oh v}=0
\,.	
\end{equation}
We will denote it as $\vo$. At this parameter value, the quark separation becomes infinite. 

Thus, at large separations the energy of the configuration is given in parametric form by $E_{\2Qqb}=E_{\2Qqb}(v)$ and $\ell=\ell(v)$. The parameter varies from $\vz$ to $\vo$. 

At this point a brief summary of our analysis is as follows. $E_{\2Qqb}$ is a piecewise function of $\ell$, and the shape of the connected configuration depends on the separation between the heavy quarks.

\subsubsection{More on the limiting cases}

Once the parametric formulas for $\ell$ and $E_{\2Qqb}$ are known, it is not difficult to analyze the behavior of $E_{\2Qqb}(\ell)$ for small and large $\ell$. This will allow us to see some important features of the model. 

We begin with the case of small $\ell$. The point is that ${\cal L}^+$ is an increasing function of $v$ which vanishes at $v=0$. Hence the limit $\ell\rightarrow 0$ makes sense only for the configuration in Figure \ref{QQqq1}, where $v$ may take a zero value. For that case, we get

\begin{equation}\label{l-small}
\ell=\sqrt{\frac{v}{\s}}\Bigl(l_0+l_1v+O(v^2)\Bigl)
\,,
\end{equation}
with $l_0=\frac{1}{2}\xi^{-\frac{1}{2}}B\bigl(\xi^2;\tfrac{3}{4},\tfrac{1}{2}\bigr)$ and $l_1=\frac{1}{2}\xi^{-\frac{3}{2}}
\bigl[ \bigl(2\xi+\frac{3}{4}\frac{\k-1}{\xi}\bigr)B\bigl(\xi^2;\tfrac{3}{4},-\tfrac{1}{2}\bigr)-B\bigl(\xi^2;\tfrac{5}{4},-\tfrac{1}{2}\bigr)\bigr]$. Here $\xi=\frac{\sqrt{3}}{2}(1-2\k-3\k^2)^{\frac{1}{2}}$ and $B(z;a,b)$ is the incomplete beta function. Similarly, the expansion for the energy is  

\begin{equation}\label{E-small}
E_{\2Qqb}=\g\sqrt{\frac{\s}{v}}\Bigl(E_0+E_1 v+O(v^2)\Bigr) +
E_{\Qqqb}+c
\,,
\end{equation}
with $E_0=1+3\k+\frac{1}{2}\xi^{\frac{1}{2}}B\bigl(\xi^2;-\tfrac{1}{4},\tfrac{1}{2}\bigr)$ and $E_1=\xi\,l_1-1-6\k+\frac{1}{2}\xi^{-\frac{1}{2}}B\bigl(\xi^2;\tfrac{1}{4},\tfrac{1}{2}\bigr)$. The constant term $E_{\Qqqb}$ is given explicitly by 

\begin{equation}\label{Qqq}
E_{\Qqqb}=\g\sqrt{\s}
\Bigl(2{\cal Q}(\bar q)-{\cal Q}(\bar v)
+2\n\frac{\ep^{\oh\bar q}}{\sqrt{\bar q}}
+3\k\frac{\ep^{-2{\bar v}}}{\sqrt{\bar v}}
\Bigr)
+c
\,.
\end{equation}
Here $\bar v$ and $\bar q$ are respectively the solutions of \eqref{vb} and \eqref{qb} in the interval $[0,1]$. In \cite{a-strb1} it was interpreted as a mass of a heavy-light antibaryon in the static limit. Because at zero baryon chemical potential the mass of $\bar Q\bar q\bar q$ coincides with that of $Qqq$, we use $E_{\Qqq}$ to refer to both masses. 

Eliminating the parameter we find

\begin{equation}\label{factor}
E_{\2Qqb}(\ell)=E_{\QQ}(\ell) + E_{\Qqq}	
\,,
\qquad
\text{with} 
\qquad
E_{\QQ}=-\frac{\alpha_{\QQ}}{\ell}+c+\boldsymbol{\sigma}_{\QQ}\ell+O(\ell^2)
\,.
\end{equation}
Here $\alpha_{\QQ}=-l_0E_0\g$ and $\boldsymbol{\sigma}_{\QQ}=\frac{1}{l_0}\Bigl(E_1+\frac{l_1}{l_0}E_0\Bigr)\g\s$. $E_{\QQ}$ is the quark-quark potential (in the antitriplet channel), and it coincides with that derived from the three quark potential in the diquark limit \cite{a-3q}. This is precisely the factorization expected from heavy quark-diquark symmetry \cite{wise}. 

We can analyze the case of large $\ell$ in a similar way. It turns out that ${\cal L}^-$ becomes infinite as $\lambda$ approaches $1$. This means that the strings in Figure \ref{QQqq4} become infinitely long. First, consider the leading approximation to the distance $\ell$ and energy $E_{\2Qqb}$. The computation is similar to those in  \cite{az1,a-QQq}. The behavior near $\lambda=1$ is given by 

\begin{equation}\label{sigular}
\ell(\lambda)=-\frac{2}{\sqrt{\s}}\ln(1-\lambda)+O(1)
\,,\qquad
\EI(\lambda)=-2\g\ep\sqrt{\s}\ln(1-\lambda)+O(1)
\,.
\end{equation}
From this, it immediately follows that  

\begin{equation}\label{Large-linear}
E_{\2Qqb}=\sigma\ell +O(1)\,,
\qquad \text{with}\qquad
\sigma=\g\ep\s
\,.	
\end{equation}
Here $\sigma$ is the physical string tension. This is one of the examples of the universality of the string tension in the model we are considering. It turns out that $\sigma$ is the same in all the cases of connected string configurations (quark-antiquark \cite{az1}, hybrid \cite{a-hyb}, three-quark \cite{a-3q0}, and $QQq$ \cite{a-QQq}). 

Next, consider $E_{\2Qqb}-\sigma\ell$. Using \eqref{l4} and \eqref{E4}, it can be written as 

\begin{equation}\label{diff}
\begin{split}
E_{\2Qqb}-\sigma\ell=&2\g\sqrt{\frac{\s}{\lambda}}
\biggl(
\int_0^1\frac{du}{u^2}
\Bigl(\ep^{\lambda u^2}
\Bigl[1-\lambda u^4\ep^{1+\lambda(1-2u^2)}\Bigr]
\Bigl[1-u^4\ep^{2\lambda(1-u^2)}\Bigr]^{-\frac{1}{2}}
-1-u^2\Bigr)\,
\\
+&
\int_{\sqrt{\frac{v}{\lambda}}}^1\frac{du}{u^2}\ep^{\lambda u^2}
\Bigl[1-\lambda u^4\ep^{1+\lambda(1-2u^2)}\Bigr]
\Bigl[1-u^4\ep^{2\lambda (1-u^2)}\Bigr]^{-\frac{1}{2}}
+
\sqrt{\frac{\lambda}{v}}
\Bigl(
\n\ep^{\oh v}+3\k\ep^{-2 v}\Bigr)\,
\biggr)+2c
\,.
\end{split}
\end{equation}
After taking the limit $v\rightarrow \vo$, we find

\begin{equation}\label{Large-constant}
E_{\2Qqb}-\sigma\ell=2\g\sqrt{\s}
\biggl(-{\cal I}(\vo)
+
\frac{\n\ep^{\oh\vo}+3\k\ep^{-2\vo}}{\sqrt{\vo}}
\biggr)+2c
\,,
\end{equation}
where the function ${\cal I}$ is defined by Eq.\eqref{I}. This can be rewritten as 

\begin{equation}\label{EQQqq-large}
	E_{\2Qqb}=\sigma\ell-2\g\sqrt{\s}\,I_{\2Qqb}+2c+o(1)
	\,,
\qquad
\text{with}
\qquad
I_{\2Qqb}={\cal I}(\vo)
-
\frac{\n\ep^{\oh v_1}+3\k\ep^{-2 v_1}}{\sqrt{v_1}}
\,.
\end{equation}
An important feature is that the constant term is different in the two expansions.

\subsection{Disconnected configurations}

\begin{figure}[htbp]
\centering
\includegraphics[width=5.65cm]{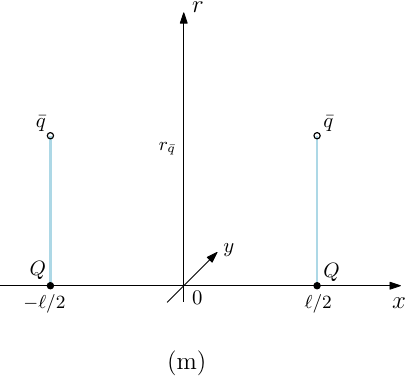}
\hspace{.3cm}
\includegraphics[width=5.64cm]{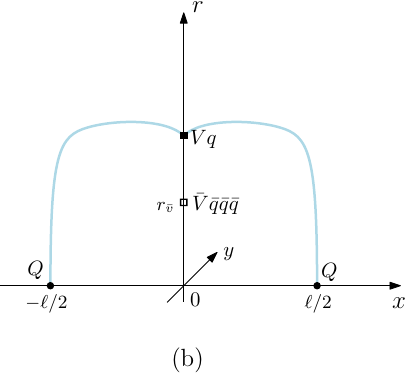}
\hspace{.3cm}
\includegraphics[width=5.64cm]{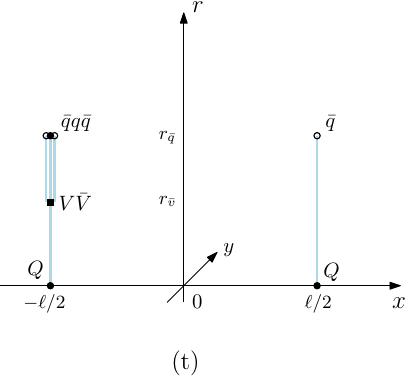}
\caption{{\small Disconnected configurations. All pairs of hadrons are non-interacting.}}
\label{disc}
\end{figure}

It is clear that in addition to the connected configuration, there may be disconnected ones. The oldest and best-known is configuration (m) in Figure \ref{disc}. It represents a pair of heavy  
light mesons. Note that unlike the $QQ\bar Q\bar Q$ system such a configuration is unique, since if the strings are connected crosswise, then the configuration is unstable. The total energy is just twice that of the meson. In the static limit the last was calculated in \cite{a-strb1} with the result 

\begin{equation}\label{Qqb}
E_{\Qqb}=\g\sqrt{\s}\Bigl({\cal Q}(\bar q)+\n \frac{\ep^{\oh \bar q}}{\sqrt{\bar q}}\Bigr)+c
\,,
\end{equation}
where $\bar q$ is a solution to Eq.\eqref{qb}. 

Now consider configuration (b). It is interpreted as a pair of baryons, one of which is a doubly heavy baryon $QQq$ and the other a light antibaryon $\bar q\bar q\bar q$. The antibaryon looks like the light antiquarks sit on top of the vertex. The total energy is the sum of two terms: $E_{\QQq}$ and $E_{\qqq}$. The former was computed in \cite{a-QQq} using the present model. On several occasions we will need the explicit formulas derived in this paper and therefore we include a brief summary of it in Appendix C. The latter was computed in \cite{a-stb3q}, but for the case of a light baryon $qqq$. Since there is no difference between these cases at zero baryon chemical potential, we have 

\begin{equation}\label{Eqqq}
	E_{\qqq}=3\g\sqrt{\frac{\s}{v_{\qqq}}}\bigl(\k\ep^{-2v_{\qqq}}+\n\ep^{\oh v_{\qqq}}\Bigr)
	\,.
\end{equation}
Here $v_{\qqq}=\s r_{\bar v}^2$ which is determined from the equation 

\begin{equation}\label{v3}
\n (1-\bar v)+\k(1+4\bar v)\ep^{-\frac{5}{2}\bar v}=0
\, 
\end{equation}
on the interval $[0,1]$. Stated simply, this equation means that the force acting on the vertex is equilibrated by that acting on the antiquarks, and as a result the antibaryon is at rest.

The last string configuration represents a pair hadrons. One of those is a heavy light tetraquark and the other a meson. The novelty here is a configuration for the tetraquark which has not been discussed before. In the static limit, it looks like the vertices and light quarks are grouped in two separate clusters one at $r=\rvb$ and one at $r=\rqb$. The total action is therefore the sum of the Nambu-Goto actions plus the actions for the vertices and background scalar. In the static gauge the action takes the form

\begin{equation}\label{actionQqqq}
	S=\g T	\biggr(
	\int_{0}^{\rvb} \frac{dr}{r^2}\,\ep^{\s r^2}\,\,
	+3\int_{\rvb}^{\rqb} \frac{dr}{r^2}\,\ep^{\s r^2}
+6\k\,\frac{\ep^{-2\s\rvb^2}}{\rvb}
+3\n\frac{\ep^{\frac{1}{2}\s\rqb^2}}{\rqb}
\,\biggr)
\,.
\end{equation}
We have evaluated the string actions on the classical solutions $x(r)=const$. The energy of the configuration is given by $E=S/T$. By virtue of Eqs.\eqref{E90} and \eqref{E|}, the energies of the straight strings can be expressed in terms of the ${\cal Q}$ function. Thus we arrive at the formula  

\begin{equation}\label{Qqqq}
E_{\Qqqq}=3\g\sqrt{\s}
\Bigl({\cal Q}(\bar q)-\frac{2}{3}{\cal Q}(\bar v)
+2\k\frac{\ep^{-2{\bar v}}}{\sqrt{\bar v}}
+\n\frac{\ep^{\oh\bar q}}{\sqrt{\bar q}}
\Bigr)
+c
\,.
\end{equation}
However, this is not the whole story as we still have to minimize the action with respect to $\rvb$ and $\rqb$. A simple calculation shows that $\bar v$ and $\bar q$ must be solutions of Eqs.\eqref{vb} and \eqref{qb}, respectively. Finally, the energy of the configuration is given by the sum of $E_{\Qqqq}$ and $E_{\Qqb}$.
 
We conclude our discussion of the disconnected configurations with a few remarks. First of all, in the string models of hadrons disconnected configurations correspond to the possible decay products of an initial bound state.\footnote{As we will discuss in the next subsection, for the $QQ\bar q\bar q$ system such a bound state can be described by the connected configuration if the separation between the heavy quarks is small enough.} In the case of interest, this implies\footnote{We restrict to the disconnected configurations which could contribute to the ground state (potential) of the $QQ\bar q\bar q$ system. Because of this, the configuration corresponding to $2Q\bar q+q\bar q$ is omitted.} 

\begin{equation}\label{decay}
\begin{split}
&\,\,Q\bar q+Q\bar q\\
\nearrow & \\
\,QQ\bar q\bar q\,\,\rightarrow & \,\,QQq+\bar q\bar q\bar q\,\\
\searrow & \\
&\,\,Qq\bar q\bar q+Q\bar q\
\,.
\end{split}
\end{equation}

One may rephrase this by saying that at large heavy quark separations the potential flattens out. Usually, such flattening is interpreted as string breaking through light quark-antiquark pair creation. In \eqref{decay} this is indeed the case for the last two decay modes, but not for the first. In stringy language that mode can be interpreted as string junction annihilation. This has a clear meaning in ten dimensions. If one identifies a string junction (baryon vertex) with a five-brane \cite{witten}, then what happens is just a brane-antibrane annihilation \cite{a-3q0}.  

Perhaps the most interesting question to ask about those modes is which mode is relevant for the ground state? To answer this question, we need to make some estimates. This is part of what we will discuss next.

\subsection{The potential}

With the formulas for the energies of the string configurations, it is straightforward now, following the same steps as in \cite{a-strb1,a-QQq} for the $Q\bar Q$ and $QQq$ systems, to obtain the potential $V_{\2Qqb}$. But before we need to specify the model parameters.  For the purposes of this paper, we will use one of the two parameter sets suggested in \cite{a-strb1}. It is mainly a result of fitting the lattice QCD data to the string model we are considering. In this case the value of $\s$ is fixed from the slope of the Regge trajectory of $\rho(n)$ mesons in the soft wall model with the geometry \eqref{metric}, and as a result, one gets $\s=0.450\,\text{GeV}^2$ \cite{a-q2}. Then, fitting the value of the string tension $\sigma$ to its value in \cite{bulava} gives $\g=0.176$. This value is smaller than the value $\g=0.196$ obtained by fitting the lattice data for the heavy quark-antiquark potential in \cite{white} but the discrepancy between these two values is not significant. The parameter $\n$ is adjusted to reproduce the lattice result for the string breaking distance in the $Q\bar Q$ system. With $\boldsymbol{\ell}_{\QQb}=1.22\,\text{fm}$ for the $u$ and $d$ quarks \cite{bulava}, that results in $\n=3.057$. In fixing the value of $\k$, one should keep in mind two things. First, the value of $\k$ can be adjusted to fit the lattice data for the three-quark potential, as is done in \cite{a-3q} for pure $SU(3)$ gauge theory. Unfortunately, at the moment, there are no lattice data available for QCD with two light quarks. Second, the range of allowed values for $\k$ is limited to $-\frac{\ep^3}{15}$ to $\km$, as dictated by Eq.\eqref{vb}. Clearly, the phenomenologically motivated value $\k=-0.102$ is out of this range as well as $\k=-0.087$ obtained from the lattice for pure gauge theory.\footnote{Note that $\k=-0.102$ is a solution to the equation $\alpha_{\QQ}=\oh\alpha_{\QQb}$ which follows from the phenomenological rule $E_{\QQ}(\ell)=\oh E_{\QQb}(\ell)$ in the limit $\ell\rightarrow 0$. See Sec.IV of \cite{a-QQq}. Another solution to this equation is $\k=-0.975$. Because its absolute value is one order of magnitude larger than the value obtained from the lattice, we discard it.} In this situation it seems natural to pick the upper bound which is most close to those. 

We are now in position to complete the discussion of the disconnected configurations. Begin with configuration (b). An important point is that $E_{\2Qqb}$ is related to $E_{\QQq}$ by the relation $E_{\2Qqb}(\ell)\approx E_{\QQq}(\ell) + E_{\Qqq} - E_{\qQb}$. The value of the discrepancy between those does not exceed $42\,\text{MeV}$, as we describe later. Using the formulas \eqref{Qqq} and \eqref{Qqb}-\eqref{Eqqq}, we can make a simple estimate: $E_{\qqq}-E_{\Qqq}+E_{\qQb}\approx 1.445\,\text{GeV}$. From this it follows that the energy $E_{\QQq}+E_{\qqq}$ is much higher than $E_{\2Qqb}$. Hence configuration (b) is irrelevant for determining the ground state of the $QQ\bar q\bar q$ system. Now consider configuration (t). In this case we estimate the difference between $E_{\Qqqq}$ and $E_{\Qqb}$. From \eqref{Qqb} and \eqref{Qqqq}, we get $E_{\Qqqq}-E_{\Qqb}\approx 1.24\,\text{GeV}$. So, the energy of configuration (t) is higher than that of (m) and we come to the conclusion that configuration (t) is irrelevant. We can summarize all this by saying that the only relevant disconnected configuration is (m).

Having understood the relevant string configurations, we can formally define the potential of the $QQ\bar q\bar q$ system: $V_{\2Qqb}=\min\Bigl(E_{\2Qqb},2E_{\Qqb}\Bigr)$. Thus it interpolates between $E_{\2Qqb}$ at small quark separations and $2 E_{\Qqb}$ at large ones. The problem with this formal definition is that it does not say precisely what happens at intermediate quark separations. A way out would be to use the same mixing analysis as in studying the phenomenon of string breaking. Following \cite{drum}, consider a model Hamiltonian of a two-state system   

\begin{equation}\label{HD}
{\cal H}(\ell)=
\begin{pmatrix}
E_{\2Qqb}(\ell) & \Theta \\
\Theta & 2E_{\Qqb} \\
\end{pmatrix}
\,.
\end{equation}
 $\Theta$ describes the strength of the mixing between the compact tetraquark state and two mesons. Then the potential is given by the smallest eigenvalue of ${\cal H}$. Explicitly, 

\begin{equation}\label{V2Qqb}
V_{\2Qqb}=\oh\Bigl(E_{\2Qqb}+2E_{\Qqb}\Bigr)
-
\sqrt{\frac{1}{4}\Bigl(E_{\2Qqb}-2E_{\Qqb}\Bigr)^2+\Theta^2}
\,.	
\end{equation}
We treat $\Theta$ as a free parameter and find its value by the best fit of our prediction to the parameterization of $V_{\2Qqb}$ suggested from the lattice studies.

\begin{figure}[htbp]
\centering
\includegraphics[width=8.88cm]{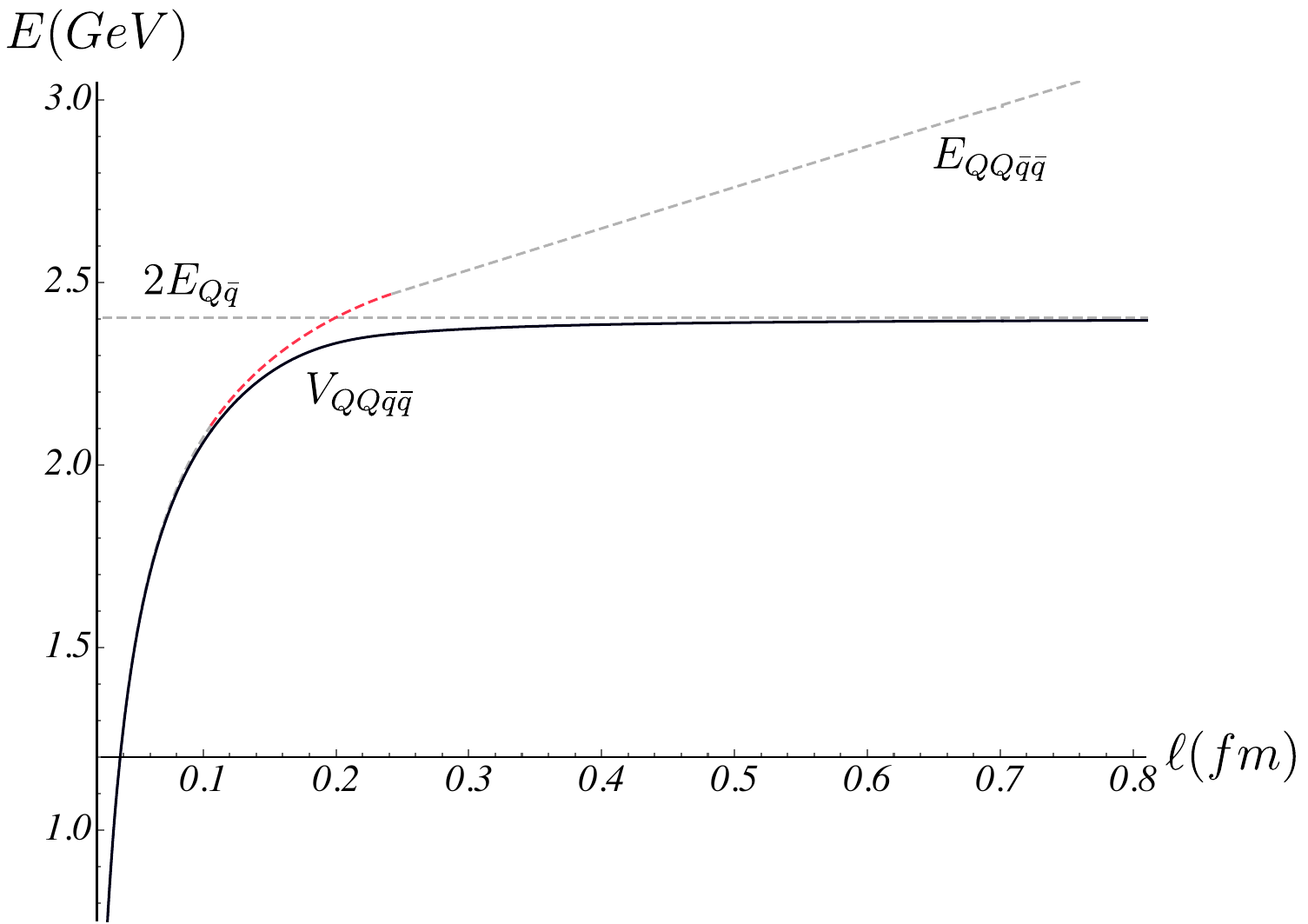}
\caption{{\small The static potential determined using the model Hamiltonian \eqref{HD}. Here and later, $c=0.623\,\text{GeV}$ and $\Theta=0.07\,\text{GeV}$. The red dashed curve corresponds to the string configuration of Figure \ref{QQqq2}.}}
\label{V}
\end{figure}

The proposal is effective, and gives a practical recipe for computing the potential of the $QQ\bar q\bar q$ system. In Figure \ref{V} we plot the potential as a function of 
quark separation. In doing so, we use the parameter values as described above. The notable feature of $V_{\2Qqb}$ is that the flattening starts at relatively small quark separations. A typical scale is of order of $0.2\,\text{fm}$, whereas that for string breaking in the $Q\bar Q$ system is of order of a $\text{fm}$. To make this more quantitative, we define a critical separation distance $\ell_{\2Qqb}$ by   

\begin{equation}\label{lc}
E_{\2Qqb}(\ell_{\2Qqb})=2E_{\Qqb}
\,.
\end{equation}
One can think of it as a scale which separates the connected and disconnected configurations, or in other words the descriptions in terms of the compact tetraquark and two mesons. Because $\ell_{\2Qqb}$ is expected to be small enough, this equation can be solved approximately by neglecting all but the first three terms in \eqref{factor}. With \eqref{Qqb}, this gives

\begin{equation}\label{lc-solution}
	\ell_{\2Qqb}
	\approx\frac{\g\sqrt{\s}}{2\boldsymbol{\sigma}_{\QQ}}\Bigl({\cal Q}(\bar v)-3\k\frac{\ep^{-2\bar v}}{\sqrt{\bar v}}\Bigr)
	+
	\sqrt{\frac{\alpha_{\QQ}}{\boldsymbol{\sigma}_{\QQ}}+\frac{\g^2\s}{4\boldsymbol{\sigma}_{\QQ}^2}\Bigl({\cal Q}(\bar v)-3\k\frac{\ep^{-2\bar v}}{\sqrt{\bar v}}\Bigr)^2}
	\,.
\end{equation}
Here $\bar v$ is a solution to Eq.\eqref{vb}. 

Let us a make a simple estimate of the critical separation distance. For the parameter values we use, we get 

\begin{equation}\label{lcnum}
\ell_{\2Qqb}\approx 0.184\,\text{fm}	
\,.
\end{equation}
Thus, this simple estimate suggests that the critical separation distance is indeed of order of $0.2\,\text{fm}$.

At this point some remarks are in order. First, $\ell_{\2Qqb}$ is finite and scheme independent. The normalization constant $c$ drops out of Eq.\eqref{lc}. Second, the solution depends on $\bar v$, which describes the position of the vertices in the bulk, and has no dependence on $\bar q$ and $\n$.\footnote{Note that  $\ell_{\2Qqb}$ is also independent of $\g$, as follows from \eqref{factor}. According to the gauge/string duality $\g$ is some function of the 't Hooft coupling.} This suggests that such defined critical separation distance is indeed related to gluonic degrees of freedom, as expected from annihilation of the baryon vertices made of gluons. Finally, $\ell_{\2Qqb}$ belongs to the range of quark separations for which the corresponding string configuration is presented in Figure \ref{QQqq2}. This configuration is the first among the three connected configurations, where the positions of the vertices coincide. Therefore it is natural to expect from string theory that the brane-antibrane annihilation occurs right here.

\section{More on the potential}
\renewcommand{\theequation}{4.\arabic{equation}}
\setcounter{equation}{0}

\subsection{The relation between $V_{\QQq}$ and $V_{\2Qqb}$}

An interesting relation can be deduced from heavy quark-diquark symmetry. Indeed, one can express $E_{\QQ}$ from Eq.\eqref{factorQQq} and then substitute it into Eq.\eqref{factor} to get 

\begin{equation}\label{quigg}
E_{\2Qqb}(\ell)=E_{\QQq}(\ell) + E_{\Qqq} - E_{\qQb}		
\,.
\end{equation}
This relation was discussed in \cite{JMR}.\footnote{See also \cite{quigg} for the relation between the corresponding hadron masses.} It is clearly true for small $\ell$. 

\begin{figure}[htbp]
\centering
\includegraphics[width=8.25cm]{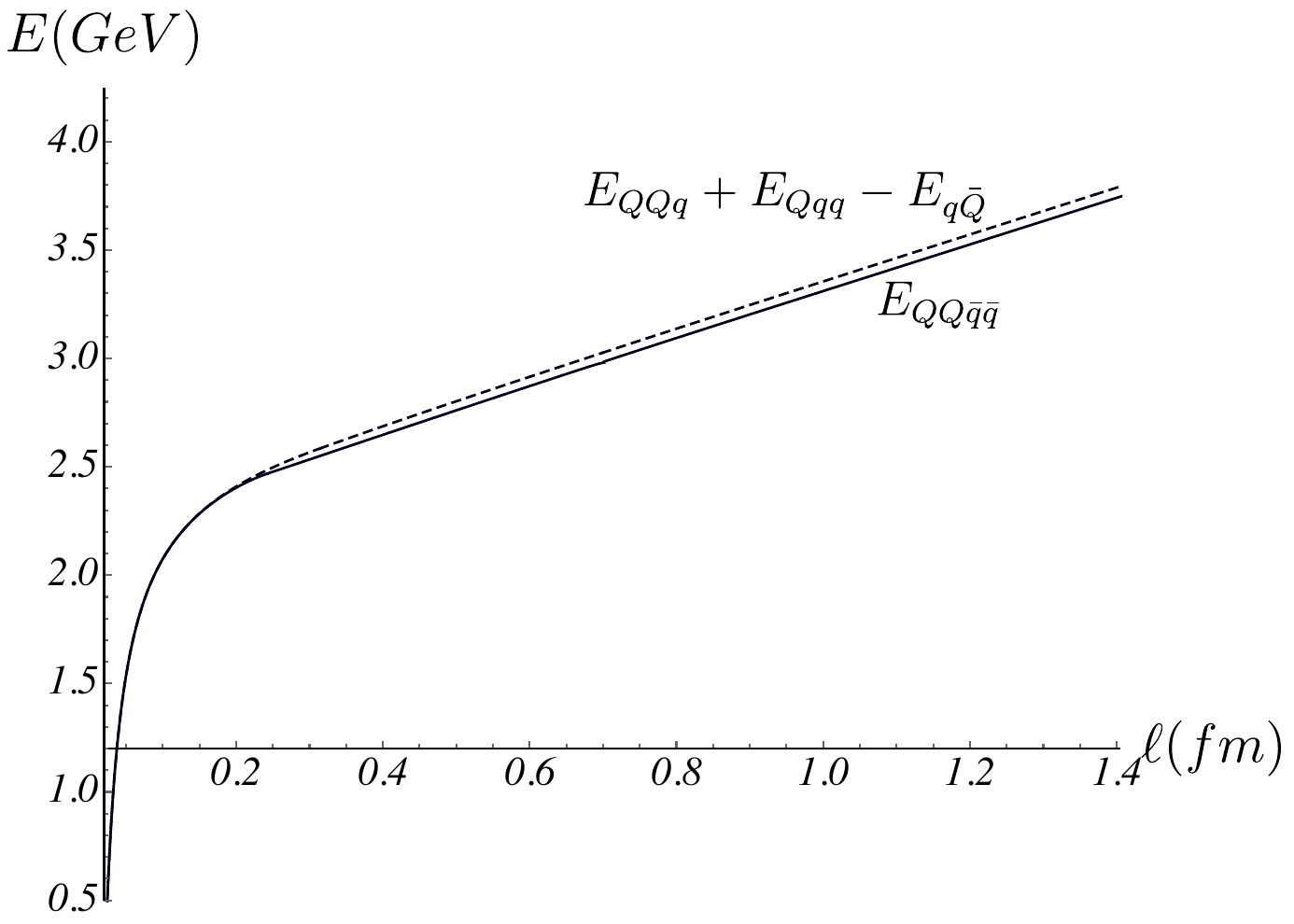}
\hspace{1.1cm}
\includegraphics[width=8.25cm]{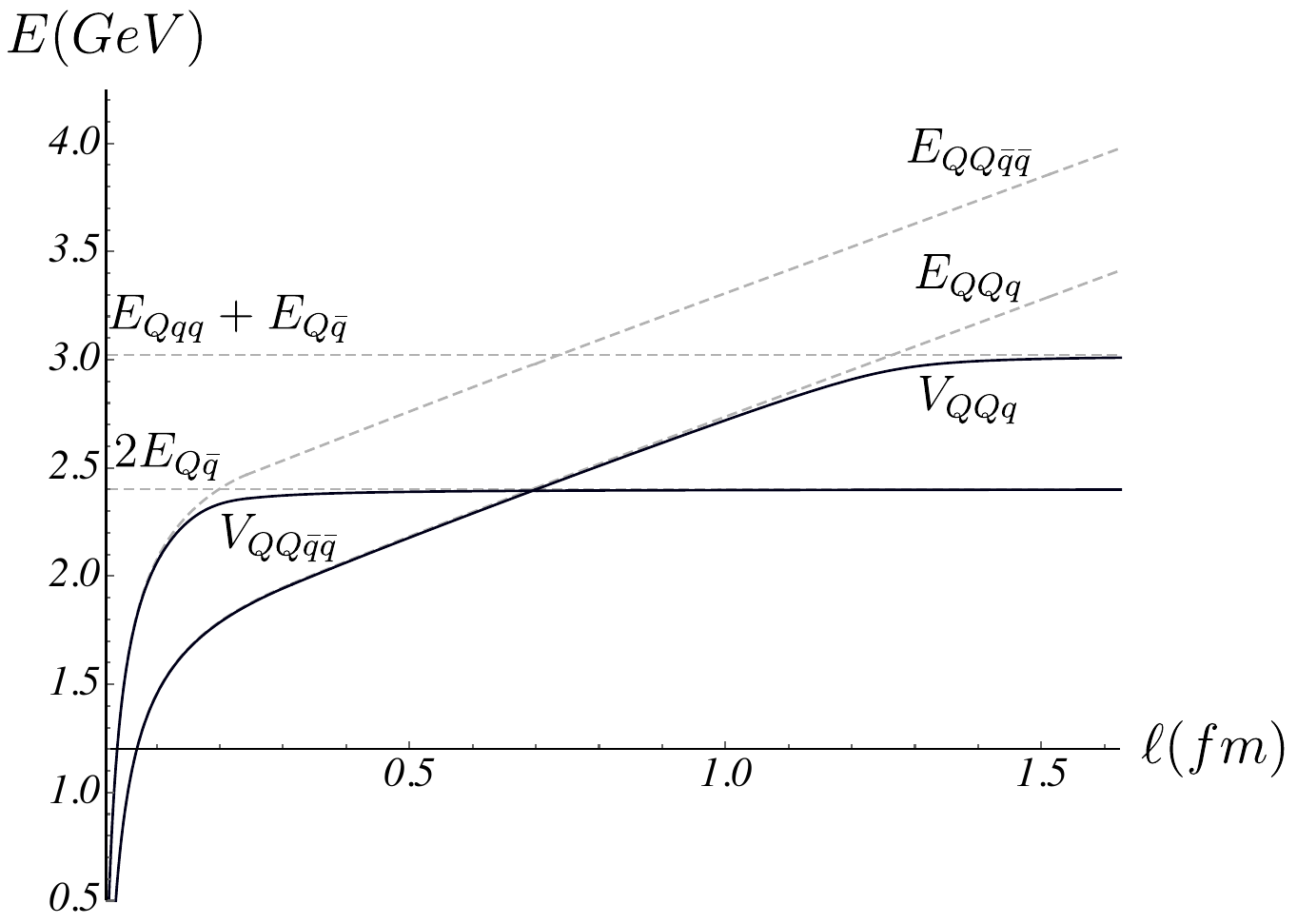}
\caption{{\small Various $E$ vs $\ell$ plots. Both potentials are described by the model Hamiltonians with $\Theta=0.07\,\text{GeV}$.}}
\label{JM}
\end{figure}

We have developed all the necessary machinery to directly check if the relation holds for large $\ell$. Using the formulas of Sect.III and Appendix C, we plot $E_{\QQq} + E_{\Qqq} - E_{\qQb}$ and $E_{\2Qqb}$ as a function of $\ell$ in Figure \ref{JM} on the left. As seen 

\noindent from the Figure, the deviation between these two curves is negligible for small separations $\ell\lesssim 0.25\,\text{fm}$, but it increases with increasing $\ell$. It is easy to find the maximum value of the deviation. Using the asymptotic expansions \eqref{EQQqq-large} and \eqref{EQQq-large}, we get

\begin{equation}\label{Gap}
\Delta=\g\sqrt{\s}\biggl[
{\cal Q}(\bar q)-{\cal Q}(\bar v)
+
2\bigl({\cal I}(\vo)-{\cal I}(\Vo)\bigr)
+
\n\biggl(\frac{\ep^{\oh\Vo}}{\sqrt{\Vo}}-2\frac{\ep^{\oh\vo}}{\sqrt{\vo}}+\frac{\ep^{\oh\bar q}}{\sqrt{\bar q}}\biggr)
+
3\k\biggl(\frac{\ep^{-2\Vo}}{\sqrt{\Vo}}-2\frac{\ep^{-2\vo}}{\sqrt{\vo}}
+\frac{\ep^{-2\bar v}}{\sqrt{\bar v}}
\biggr)	
\biggl]
\,.
\end{equation}
Here $\bar q$ is a solution to \eqref{qb}, $\bar v$ to \eqref{vb}, $\vo$ to \eqref{v1}, and $\Vo$ to \eqref{v1-QQq}. It is interesting to make an estimate of $\Delta$. For the parameter values we are using, the calculation gives $\Delta\approx 42\,\text{MeV}$. Thus, the relation \eqref{quigg} seems quite acceptable for phenomenological purposes. 

However, the above conclusion does not remain valid for the corresponding potentials. This is because of the flattening of the potentials happens at two different scales. For $V_{\2Qqb}$ the critical separation is of order $0.184\,\text{fm}$, whereas for $V_{\QQq}$ of order of $1.257\,\text{fm}$.\footnote{ The latter is the result of estimating the string breaking distance \eqref{lcQQq} for the parameter values as above \cite{a-QQq}.} In Figure \ref{JM} on the right, we plot the potentials to illustrate this effect. The physical reason for such a big difference between the scales is the different nature of the flattening in the two cases. In the first case it is associated to the vertex annihilation process, while in the second to the string breaking phenomenon by light quark pair production. 

We conclude by giving the refined version of \eqref{quigg}

\begin{equation}\label{oda}
V_{\2Qqb}(\ell)=V_{\QQq}(\ell) + E_{\Qqq} - E_{\qQb}	
	\,,\qquad
	\text{if}
	\qquad \ell\lesssim 0.2\,\text{fm}
	\,.
\end{equation}

\subsection{Comparison with the lattice}

The potentials for the $QQ\bar q\bar q$ system have been studied on the lattice \cite{wagner}. In that case those are extracted from the correlators of meson operators. The results are consistently parameterized by 

\begin{equation}\label{MW}
V_{\2Qqb}(\ell)=-\frac{\alpha}{\ell}\exp\Bigl(-\frac{\ell^{\,p}}{d^{\, p}}\,\Bigl)\,+2 E_{\Qqb}
\,,
\end{equation}
with parameters $\alpha$, $d$ and $p$. 

To make contact with the results of Sect.III, we use the small $\ell$ expansion \eqref{factor} and solve  for the unknown coefficients, with the result

\begin{equation}\label{dp}
	\alpha=\alpha_{\QQ}
	\,,\qquad
	d=\sqrt{\frac{\alpha_{\QQ}}{\boldsymbol{\sigma}_{\QQ}}}
	\,,\qquad
	p=2
	\,.
\end{equation}
The parameters are described in terms of the coefficients of the quark-quark potential, as one should expect from heavy quark-diquark symmetry. 

To go further, let us make a simple estimate of the screening length $d$. For the parameter values of Sec.III, $d=0.20\,\text{fm}$.  It is worth noting that the contribution of the first term in \eqref{lc-solution} turns out to be of order $-0.017\,\text{fm}$. This explains the small difference between the values of $d$ and $\ell_{\2Qqb}$. Both our estimates are inside the range of the ones found on the lattice for isospin one states \cite{wagner}, $d=0.16^{+0.05}_{-0.02}\,\text{fm}$. So at this point the agreement with the lattice results is good. 

The problem arises when matching the constant terms in \eqref{factor} and \eqref{MW}. In general, $2E_{\Qqb}\not =E_{\Qqq}+c$, unless the parameters are adjusted. Note that $c$ drops out of the inequality, as it should be with an infinite quark mass. To get around this problem, one should consider instead another function in \eqref{MW} having a non-zero constant term in its expansion for small $\ell$. This would ensure the possibility for the factorization.

\begin{figure}[htbp]
\centering
\includegraphics[width=8.25cm]{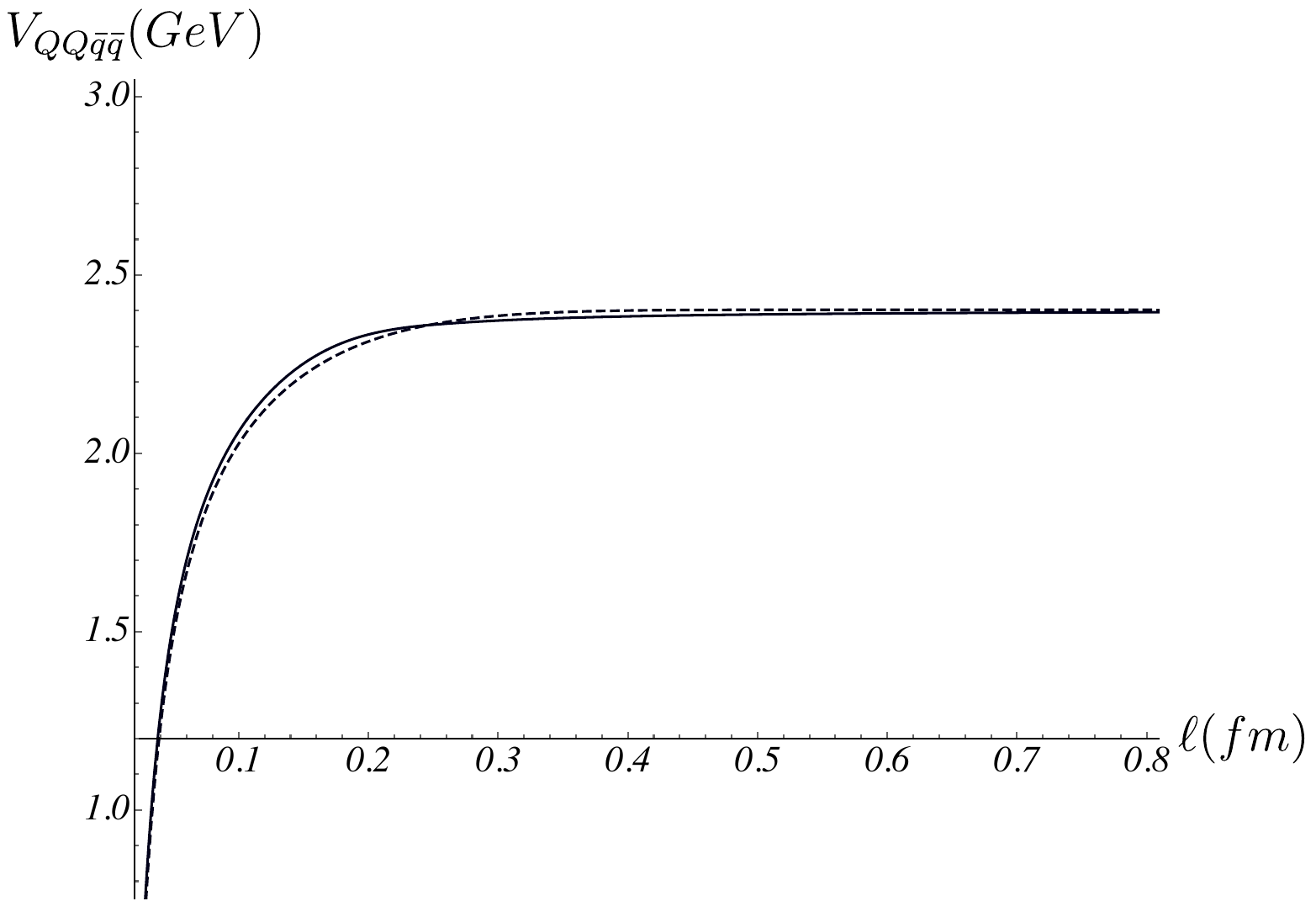}
\caption{{\small $V_{\2Qqb}$ vs $\ell$. The solid curve corresponds to the string model of Sec.III, whereas the dashed and dotted curves to the parameterization \eqref{MW} with $\k=-\frac{1}{4}\ep^{\frac{1}{4}}$ and $\k=-0.102$.}}
\label{VMW}
\end{figure}

Nevertheless it is interesting to compare the result of Sect.III to the parameterization \eqref{MW}. In Figure \ref{VMW} we draw the graphs for both cases, and we use those to fix the value of $\Theta$. The latter gives $\Theta=0.07\,\text{GeV}$. There are two things to  
be mentioned about this. The first is a visible deviation between the solid and dashed curves on the interval $0.08\,\text{fm}\lesssim \ell\lesssim 0.22\,\text{fm}$. The appearance of this deviation can be ascribed to the mismatch of the constant terms in the small $\ell$ expansions, and its value is of order of $41\,\text{MeV}$ as follows from  $E_{\Qqq}+c-2E_{\Qqb}\approx 41\,\text{ MeV}$. The second is a falloff at large $\ell$. It is power-law for \eqref{V2Qqb}, but exponential for \eqref{MW}. The reason for the power-law falloff is the choice of $\Theta=const$. In fact, one can get the exponential falloff by taking $\Theta$ as a Gaussian function with its peak at $\ell=\ell_{\2Qqb}$, as done for example in the case of the $Q\bar Q$ system \cite{gon} . If so, then one additional parameter is required, the Gaussian width. 

As noted above, one of the limitations of the construction is that the phenomenologically motivated value of $\k$ is out of the allowed range. The question arises of whether our conclusions on the properties of $V_{\2Qqb}$ hold also for this case. The parameterization \eqref{MW} being well defined at $\k=-0.102$ may help to shed some light on this question. First, let us estimate the screening length. Using \eqref{dp}, we get $d=0.17\,\text{fm}$. The estimate suggests that the screening length at $\k=-0.102$ is smaller than at $\k=-\frac{1}{4}\ep^{\frac{1}{4}}$, but still inside the range found on the lattice. Next we plot $V_{\2Qqb}$ versus $\ell$. As seen in Figure \ref{VMW}, the change in $\k$ shifts the plot to the left, but without any essential differences amongst each other. This provides rather strong evidence that the outcome is robust to this change in $\k$. 
\section{Concluding Comments}
\renewcommand{\theequation}{5.\arabic{equation}}
\setcounter{equation}{0}

(i) What we have learned from the string models is that the two potentials $V_{\2Qqb}$ and $V_{\QQq}$ look very similar except for one crucial difference: they get flattened at the well separated scales. This implies that the latter being more deep could have more bound (excited) states. If so, then the relation between the masses of the ground-state hadrons \cite{quigg} is no longer valid for the excited hadrons. Hopefully, it will be possible eventually to check this prediction by computer simulations.  

\begin{figure}[htbp]
\centering 
\includegraphics[width=3.8cm]{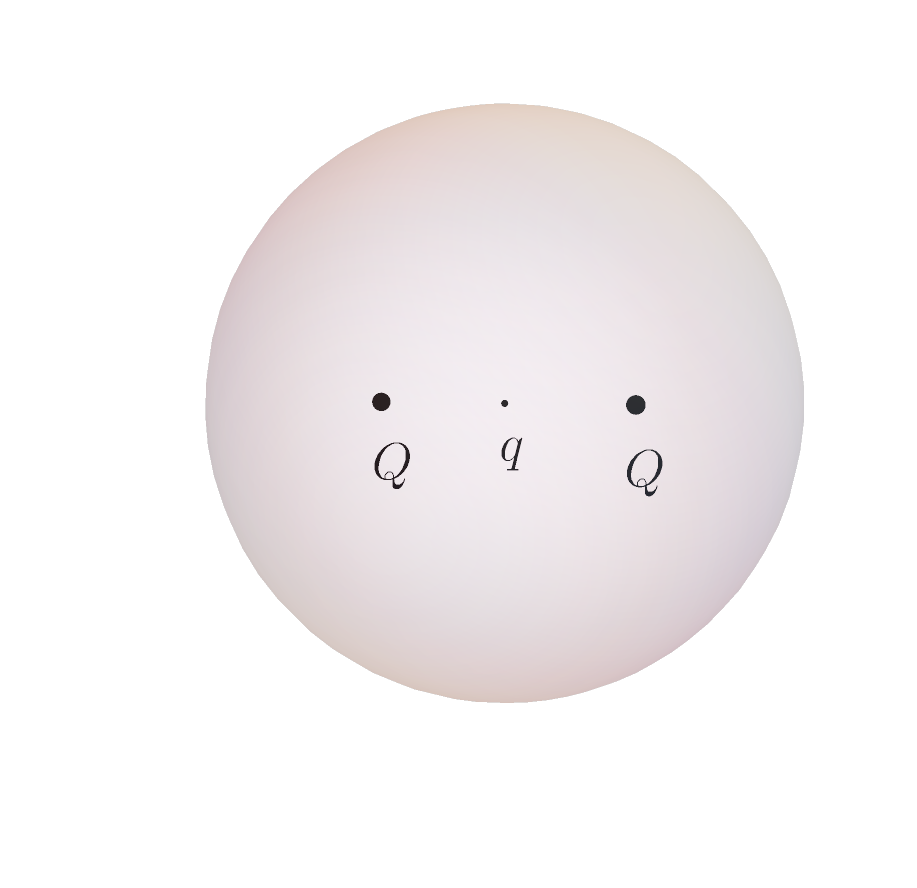}
\hspace{3cm}
\includegraphics[width=3.8cm]{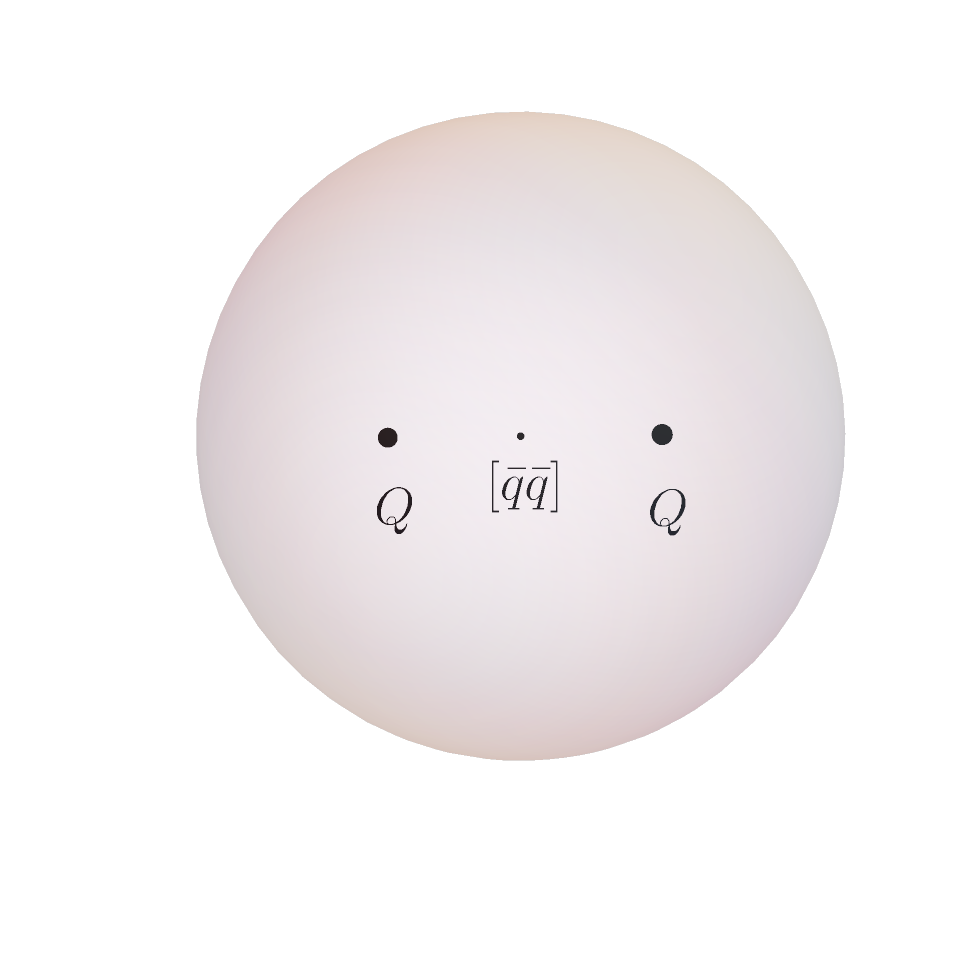}
\caption{{\small The hadro-quarkonium pictures for the $QQq$ and $QQ\bar q\bar q$ systems at small separation between the heavy quarks. The heavy quarks are embedded in the clouds of the light quark and antidiquark.}}
\label{V-small}
\end{figure}

(ii) The string theory argument leading to the relation between the potentials $V_{\2Qqb}$ and $V_{\QQq}$ at small quark separations is the similarity of the connected string configurations of Figures \ref{QQqq1} and \ref{cQQq}. From the four-dimensional point of view, the light quark is in the middle between the heavy quarks of the $QQq$ system, whereas an antidiquark is formed in the middle between the quarks of the $QQ\bar q\bar q$ system. These are the cases sketched in Figure \ref{V-small}. Note that     

\noindent both are in line with the symmetry expectations on the ground states, as the most symmetric spatial configurations of the quarks.

(iii) We have treated the off-diagonal element $\Theta$ of the model Hamiltonian as a free parameter. It is of further interest to develop a string theory technique which would allow a direct computation of it.  

(iv) The string theory approach allows us to naturally define a scale, called the critical separation distance, below which the $QQ\bar q\bar q$ system can be thought of mainly as a compact tetraquark and above as a pair of mesons. A simple estimate gives

\begin{equation}\label{lcratio}
\frac{\ell_{\QQb}}{\ell_{\2Qqb}}\approx 6.630
\,.
\end{equation}
So, the critical separation distance is much smaller than the string breaking distance in the $Q\bar Q$ system. This might be helpful in the interpretation of exotic mesons containing two heavy quarks and two light antiquarks, and vice versa.

(v) Our analysis is also applicable for the $\bar Q\bar Q qq$ system. The corresponding string configurations are obtained by replacing $Q\rightarrow \bar Q$, $\bar q\rightarrow q$, $V\rightarrow \bar V$, and $\bar V\rightarrow V$.

\begin{acknowledgments}
We are grateful to J.-M. Richard for helpful comments, and to M. Wagner for communications. This research is supported by Russian Science Foundation grant 20-12-00200 in association with Steklov Mathematical Institute.
\end{acknowledgments}

\appendix
\section{Notation and definitions}
\renewcommand{\theequation}{A.\arabic{equation}}
\setcounter{equation}{0}
In all Figures throughout the paper, heavy and light quarks (antiquarks) are denoted by $Q$ and $q\,(\bar q)$, and baryon (antibaryon) vertices by $V\,(\bar V)$. We assume that all strings are in the ground state. So, these strings are represented by curves without cusps, loops, etc. When not otherwise noted, we usually set light quarks (antiquarks) at $r=\rq\,(\rqb)$ and vertices at $r=\rv\,(\rvb)$. For convenience, we introduce dimensionless variables: $q=\s\rq^2$, $\bar q=\s\rqb^2$, $v=\s\rv^2$, and $\bar v=\s\rvb^2$. They take values on the interval $[0,1]$ and show how far from the soft-wall these objects are.\footnote{In these dimensionless units, the soft wall is located at $1$.} 

In order to write formulas briefly, we use the set of basic functions \cite{a-stb3q}: 

\begin{equation}\label{fL+}
{\cal L}^+(\alpha,x)=\cos\alpha\sqrt{x}\int^1_0 du\, u^2\, \ep^{x (1-u^2)}
\Bigl[1-\cos^2{}\hspace{-1mm}\alpha\, u^4\ep^{2x(1-u^2)}\Bigr]^{-\frac{1}{2}}
\,,
\qquad
0\leq\alpha\leq\frac{\pi}{2}\,,
\qquad 
0\leq x\leq 1
\,.
\end{equation}
It is a non-negative function which vanishes if $\alpha=\frac{\pi}{2}$ or $x=0$, and has a singular point at $(0,1)$;

\begin{equation}\label{fL-}
{\cal L}^-(y,x)=\sqrt{y}
\biggl(\,
\int^1_0 du\, u^2\, \ep^{y(1-u^2)}
\Bigl[1-u^4\,\ep^{2y(1-u^2)}\Bigr]^{-\frac{1}{2}}
+
\int^1_
{\sqrt{\frac{x}{y}}} 
du\, u^2\, \ep^{y(1-u^2)}
\Bigl[1-u^4\,\ep^{2y(1-u^2)}\Bigr]^{-\frac{1}{2}}
\,\biggr)
\,,
\quad
0\leq x\leq y\leq 1
\,,
\end{equation}
which is also a non-negative. It vanishes at the origin and becomes singular at $y=1$. At $y=x$, ${\cal L}^-$ reduces to ${\cal L}^+$ with $\alpha=0$;  

\begin{equation}\label{fE+}
{\cal E}^+(\alpha,x)=\frac{1}{\sqrt{x}}
\int^1_0\,\frac{du}{u^2}\,\biggl(\ep^{x u^2}
\Bigl[
1-\cos^2{}\hspace{-1mm}\alpha\,u^4\ep^{2x (1-u^2)}
\Bigr]^{-\frac{1}{2}}-1-u^2\biggr)
\,,
\qquad
0\leq\alpha\leq\frac{\pi}{2}\,,
\qquad 
0\leq x\leq 1
\,.
\end{equation}
This function is singular at $x=0$ and $(0,1)$;

\begin{equation}\label{fE-}
{\cal E}^-(y,x)=\frac{1}{\sqrt{y}}
\biggl(
\int^1_0\,\frac{du}{u^2}\,
\Bigl(\ep^{y u^2}\Bigl[1-u^4\,\ep^{2y(1-u^2)}\Bigr]^{-\frac{1}{2}}
-1-u^2\Bigr)
+
\int^1_{\sqrt{\frac{x}{y}}}\,\frac{du}{u^2}\,\ep^{y u^2}
\Bigl[1-u^4\,\ep^{2y(1-u^2)}\Bigr]^{-\frac{1}{2}}
\biggr) 
\,,
\,\,\,
0\leq x\leq y\leq 1
\,.
\end{equation}
It is singular at $(0,0)$ and $y=1$. Just like for the ${\cal L}$'s, ${\cal E}^-$ reduces to ${\cal E}^+$ at $y=x$;

\begin{equation}\label{Q}
{\cal Q}(x)=\sqrt{\pi}\text{erfi}(\sqrt{x})-\frac{\ep^x}{\sqrt{x}}
\,.
\end{equation}
Here $\text{erfi}(x)$ is the imaginary error function. ${\cal Q}$ is the special case of ${\cal E}^+$ obtained by setting $\alpha=\frac{\pi}{2}$. A useful fact is that its small $x$ behavior is 

\begin{equation}\label{Q0}
{\cal Q}(x)=-\frac{1}{\sqrt{x}}+\sqrt{x}+O(x^{\frac{3}{2}})
\,;
\end{equation}

\begin{equation}\label{I}
	{\cal I}(x)=\int_0^1\frac{du}{u^2}\Bigl(1+u^2-\ep^{u^2}\Bigl[1-u^4\ep^{2(1-u^2)}\Bigr]^{\frac{1}{2}}\Bigr)
-
\int_{\sqrt{x}}^1\frac{du}{u^2}\ep^{u^2}\Bigl[1-u^4\ep^{2(1-u^2)}\Bigr]^{\frac{1}{2}}\,
\,,
\qquad
0< x\leq 1
\,.
\end{equation}
In fact, the first integral can be evaluated numerically with the result $0.751$.

\section{A static Nambu-Goto string with fixed endpoints}
\renewcommand{\theequation}{B.\arabic{equation}}
\setcounter{equation}{0}

The purpose of this appendix is to briefly review some facts about a static Nambu-Goto string in the curved geometry \eqref{metric} that are helpful for understanding the string configurations of Sect.III. For more details on those facts, see \cite{a-3qPRD, a-stb3q}. 

\begin{figure}[htbp]
\centering
\includegraphics[width=4.3cm]{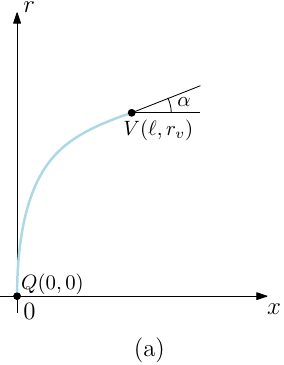}
\hspace{2cm}
\includegraphics[width=4.3cm]{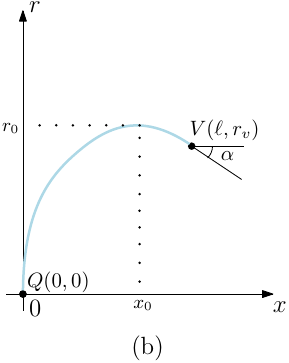}
\hspace{2cm}
\includegraphics[width=3.6cm]{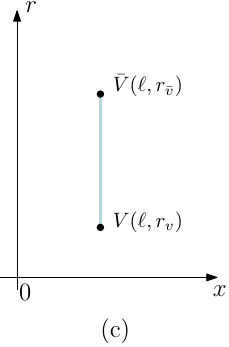}
\caption{{\small A static string stretched between two points. $\alpha$ is the tangent angle. (a) $0<\alpha<\frac{\pi}{2}$. (b) $-\frac{\pi}{2}<\alpha<0$. A turning point is at $x=x_0$. (c) A special case $\alpha=\frac{\pi}{2}$.}}
\label{ng}
\end{figure}

For our purposes, the only cases we need to consider are presented in Figure \ref{ng}.

In case (a), the string length along the $x$-axis and its energy can be written as 

\begin{equation}\label{lE+}
\ell=\frac{1}{\sqrt{\s}}{\cal L}^+(\alpha,v)
\,,\qquad
E=\g\sqrt{\s}\,{\cal E}^+(\alpha,v)+c
\,.
\end{equation}
Here $v$ is defined by $v=\s\rv^2$. $c$ is a constant arising from the use of the non-minimal subtraction scheme. If $\alpha=\frac{\pi}{2}$, then 

\begin{equation}\label{E90}
E=\g\sqrt{\s}{\cal Q}(v)+c
\,.	
\end{equation}
This is a special case of the string stretched along the $r$-axis.

In case (b), the corresponding formulas are 

\begin{equation}\label{lE-}
\ell=
\frac{1}{\sqrt{\s}}{\cal L}^-(\lambda,v)
\,,\qquad
E
= 
\g\sqrt{\s}\,{\cal E}^-(\lambda,v)+c
\,,
\end{equation}
with $\lambda=\s r_0^2$ such that $v<\lambda$. $c$ is the same normalization constant as before. Importantly, $\lambda$ is a function of $v$ and $\alpha$ of the form 

\begin{equation}\label{lambda}
\lambda=-\text{ProductLog}\bigl(-v\ep^{-v}/\cos\alpha\bigr)
\,.	
\end{equation}
Here $\text{ProductLog}(z)$ is the principal solution for $w$ in $z=w\,\ep^w$ \cite{wolfram}.

Finally, in (c) the string energy is given by 

\begin{equation}\label{E|}
E=\g\sqrt{\s}\bigl({\cal Q}(\bar v)-{\cal Q}(v)\bigr)
\,,
\end{equation}
with $\bar v=\s\rvb^2$. 

\section{The $QQq$ system}
\begin{figure}[htbp]
\centering
\includegraphics[width=5.6cm]{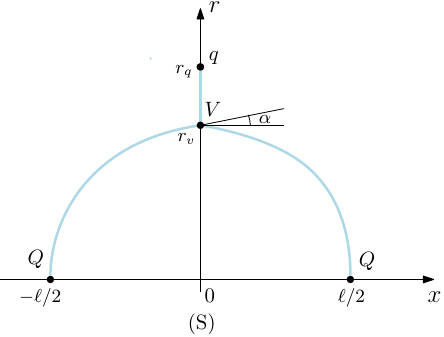}
\hspace{0.2cm}
\includegraphics[width=5.6cm]{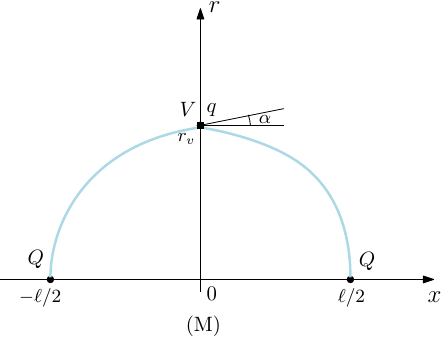}
\hspace{0.2cm}
\includegraphics[width=5.8cm]{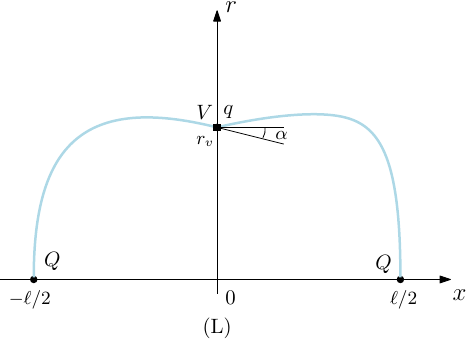}
\caption{{\small Three types of static string configurations that contribute to the potential of the $QQq$ system. $\alpha$ denotes the tangent angle of the left string.}}
\label{cQQq}
\end{figure}

\renewcommand{\theequation}{C.\arabic{equation}}
\setcounter{equation}{0}
Now let us briefly review the string construction for the $QQq$ system proposed in \cite{a-QQq}, whose conventions we follow here. First consider the connected configurations of Figure \ref{cQQq}. 
 The important point here is that the shape of the configuration changes with the increase of heavy quark separation. 

For small $\ell$ the corresponding configuration is labeled by (S). In this 
case, the total action is the sum of the Nambu-Goto actions plus the actions for the baryon vertex and background scalar. The relation between the energy and heavy quark separation is written in parametric form  

\begin{equation}\label{EQQqs}
\ell=\frac{2}{\sqrt{\s}}{\cal L}^+(\alpha,v)
\,,\qquad
E_{\QQq}=\g\sqrt{\s}
\Bigl(
2{\cal E}^+(\alpha,v)
+
\n\frac{\ep^{\oh q}}{\sqrt{q}}+3\k\frac{\ep^{-2v}}{\sqrt{v}}
+
{\cal Q}(q)-{\cal Q}(v)
\Bigr)
+2c\,,
\end{equation}
with the parameter $v$ varying from $0$ to $q$. The functions ${\cal L}^+$ and ${\cal E}^+$ are as defined in Appendix A. The value of $q$ is determined from equation 

\begin{equation}\label{q}
\ep^{\frac{q}{2}}+\n(q-1)=0	
	\,,
\end{equation} 
which is nothing else but the force balance equation at $r=\rq$. $c$ is a normalization constant. The tangent angle $\alpha$ can be expressed in terms of $v$ by using the force balance equation at $r=\rv$, with the result

\begin{equation}\label{alpha1}
\sin\alpha=\oh\bigl(1+3\k(1+4v)\ep^{-3v}\bigr)
\,.	
\end{equation}

An important point is that in the limit $\ell\rightarrow 0$ the energy reduces to a sum of energies: 

\begin{equation}\label{factorQQq}
E_{\QQq}(\ell)=E_{\QQ}(\ell)+E_{\qQb}
\,,
\end{equation}
as expected from heavy quark-diquark symmetry. The first term corresponds to the heavy quark-quark potential whose explicit form is given by Eq.\eqref{factor}, and the second to the rest energy of a heavy-light meson in the static limit. Explicitly, it is given by the same formula \eqref{Qqb} as that for $Q\bar q$. 

For intermediate values of $\ell$, the configuration is labeled by (M). It differs from the first by the absence of the string stretched between the vertex and light quark so that the quark sits on top of the vertex. So, the distance $\ell$ is expressed in terms of $v$ and $\alpha$ by the same formula as before, only for another parameter range, whereas the energy by 

\begin{equation}\label{EQQqm} 
E_{\QQq}=\g\sqrt{\s}
\Bigl(
2{\cal E}^+(\alpha,v)
+
\frac{1}{{\sqrt{v}}}\bigl(
\n\ep^{\oh v}+
3\k\ep^{-2v}
\bigr)
\Bigr)
+2c\,.
\end{equation}
The parameter $v$ varies from $q$ to $\Vz$, where $\Vz$ is a solution to 

\begin{equation}\label{v0-QQq}
\n (1-v)+3\k(1+4v)\ep^{-\frac{5}{2}v}=0
\,.
\end{equation}

The force balance equation at $r=\rv$ becomes  

\begin{equation}\label{alpha2b}
\sin\alpha=\oh\bigl(
\n (1-v)\ep^{-\oh v}+3\k(1+4v)\ep^{-3v}
\bigr)
\,.
\end{equation}
A noteworthy fact is that $\alpha(\Vz)=0$. 

For large $\ell$, the proper configuration is that labeled by (L). In fact, what happens in the transition from (M) to (L) is that the tangent angle changes the sign from positive to negative. Keeping this in mind makes it much easier to arrive at the desired result. After replacing ${\cal L}^+$ and ${\cal E}^+$ by ${\cal L}^-$ and ${\cal E}^-$, one gets

\begin{equation}\label{EQQql}
\ell=\frac{2}{\sqrt{\s}}
{\cal L}^-(\lambda,v)
\,,
\qquad
E_{\QQq}=\g\sqrt{\s}
\Bigl(
2{\cal E}^-(\lambda,v)
+
\frac{1}{\sqrt{v}}
\bigl(\n\ep^{\oh v}+3\k\ep^{-2v}\bigr)
\Bigr)
+2c\,,
\end{equation}
with the parameter $v$ varying from $\Vz$ to $\Vo$. The upper bound is found by solving the non-linear equation 

\begin{equation}\label{v1-QQq}
2\sqrt{1-v^2\ep^{2(1-v)}}+3\k(1+4v)\ep^{-3v}+\n (1-v)\ep^{-\oh v}=0
\,
\end{equation}
on the interval $[0,1]$. Using \eqref{lambda}, $\lambda$ can be expressed in terms of $v$ 

\begin{equation}\label{lambda-QQq}
\lambda(v)=-\text{ProductLog}
\biggl(-v\ep^{-v}
\Bigl(1-\frac{1}{4}\Bigl(3\k(1+4v)\ep^{-3v}
+
\n(1-v)\ep^{-\oh v}\Bigr)^2
\,\Bigr)^{-\frac{1}{2}}
\,\biggr)
\,.
\end{equation}
Note that $\lambda(\Vo)=1$ that corresponds to the limit of infinitely long strings. 

For future reference, it is worth noting that the asymptotic behavior of $E_{\QQq}(\ell)$ for large $\ell$ is 

\begin{equation}\label{EQQq-large}
	E_{\QQq}=\sigma\ell-2\g\sqrt{\s}I_{\QQq}+2c+o(1)
	\,,
\qquad
\text{with} 
\qquad
	I_{\QQq}={\cal I}(\Vo)
	-
\frac{\n\ep^{\oh\Vo}+3\k\ep^{-2 \Vo}}{2\sqrt{\Vo}}
\,
\end{equation}
and the same string tension $\sigma$ as in \eqref{Large-linear}. The function ${\cal I}$ is defined in Appendix A.

One can summarize all this by saying that the energy of the connected configuration as a function of the heavy quark separation is given in parametrical form by the two piecewise functions $E_{\QQq}=E_{\QQq}(v)$ and $\ell=\ell(v)$.

\begin{figure}[htbp]
\centering
\includegraphics[width=5.75cm]{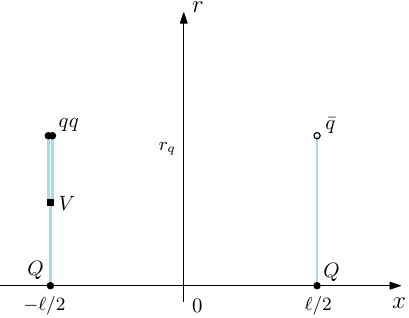}
\caption{{\small A pair of non-interacting hadrons. At zero baryon chemical potential, $\rq=\rqb$.}}
\label{dQQq}
\end{figure}

For the doubly heavy baryon $QQq$ the dominant decay mode via string breaking is 

\begin{equation}\label{decayQQq}
QQq\rightarrow Qqq\,+\,Q\bar q
\,.	
\end{equation}
In the $5d$ string models, one can think of the decay products as sketched in Figure \ref{dQQq} \cite{a-strb1}.
From \eqref{Qqq} and \eqref{Qqb}, it follows that the rest energy of the pair is 

\begin{equation}\label{E-dis}
E_{\Qqq}+E_{\Qqb}=3\g\sqrt{\s}
\biggl({\cal Q}(q)-\frac{1}{3}{\cal Q}(v)+\n\frac{\ep^{\oh q}}{\sqrt{q}}
+\k\frac{\ep^{-2v}}{\sqrt{v}}\,
\biggr)+2c
\,.
\end{equation}
Here $q$ and $v$ are the solutions to Eqs.\eqref{q} and \eqref{vb}, respectively. 

The string breaking distance is a natural scale that characterizes this decay. It is defined by equating the energies of the two configurations

\begin{equation}\label{lc-QQq}
E_{\QQq}(\boldsymbol{\ell}_{\QQq})=E_{\Qqq}+E_{\Qqb} 
\,.	
\end{equation}
The equation simplifies drastically at large $\ell$, where the phenomenon of string breaking is expected to occur. Combining \eqref{EQQq-large} and \eqref{E-dis} leads to

\begin{equation}\label{lcQQq}
\boldsymbol{\ell}_{\QQq} =\frac{3}{\ep\sqrt{\s}}
\biggl(
{\cal Q}(q)-\frac{1}{3}{\cal Q}(v) +\n\frac{\ep^{\oh q}}{\sqrt{q}}+\k\frac{\ep^{-2v}}{\sqrt{v}}+\frac{2}{3}{\cal I}_{\QQq}
\biggr)
\,.
\end{equation}
The resulting expression is independent of $c$ as it should be.

The potential $V_{\QQq}$ is given by the smallest eigenvalue of a model Hamiltonian 

\begin{equation}\label{HD-QQq}
{\cal H}(\ell)=
\begin{pmatrix}
E_{\QQq}(\ell) & \Theta' \\
\Theta' & E_{\Qqq}+E_{\Qqb} \\
\end{pmatrix}
\,,
\end{equation}
with $\Theta'$ describing the mixing between the two states. 



\small
\begin{thebibliography}{99}
\bibitem{GM}
M. Gell-Mann, Phys.Lett. {\bf 8}, 214 (1964).
\bibitem{zweig}
G. Zweig, An $SU(3)$ model for strong interaction symmetry and its breaking, CERN preprint 8182/TH.401.
\bibitem{ali}
A. Ali, L. Maiani, and A.D. Polosa, Multiquark Hadrons, Cambridge University Press, 2019.
\bibitem{LHCb}
R. Aaij et al. [LHCb], Observation of an exotic narrow doubly charmed tetraquark, arXiv:2109.01038 [hep-ex].
\bibitem{JMR0} It has a long history, going back to J.P. Ader, J.M. Richard, and P. Taxil, Phys.Rev.D {\bf 25}, 2370 (1982). For more details and references, see J.M. Richard, Doubly Heavy Tetraquarks: Lessons from Atomic Physics, a talk at "{\it LHCb mini-workshop}", CERN, September 2021. 
\bibitem{AF}
A. Francis and R. Lewis, $T_{cc}$ and its heavier cousins: Structure and binding in lattice QCD, a talk at "{\it LHCb mini-workshop}", CERN, September 2021.
\bibitem{book-u}
J. Casalderrey-Solana, H. Liu, D. Mateos, K. Rajagopal, and U.A. Wiedemann, Gauge/String Duality, Hot QCD and Heavy Ion Collisions, Cambridge University Press, 2014.
\bibitem{a-3q0}
O. Andreev, Phys.Rev.D {\bf 78}, 065007 (2008).
\bibitem{sw}
J. Sonnenschein and D. Weissman, Nucl.Phys.B {\bf 920}, 319 (2017). 
\bibitem{voloshin}
M.B. Voloshin, Deciphering the XYZ States, a talk at "{\it 17th Conference on Flavor Physics and CP Violation (FPCP 2019)}, arXiv:1905.13156 [hep-ph].
\bibitem{a-strb1}
O. Andreev, Phys.Lett.B {\bf 804} (2020) 135406.
\bibitem{az1}
O. Andreev and V.I. Zakharov, Phys.Rev.D {\bf 74}, 025023 (2006).
\bibitem{a-hyb}
O. Andreev, Phys.Rev.D {\bf 86}, 065013 (2012).
\bibitem{a-3qPRD} 
 O. Andreev, Phys.Rev.D {\bf 93}, 105014 (2016).
\bibitem{son}
J. Erlich, E. Katz, D.T. Son, and M.A. Stephanov, Phys.Rev.Lett. {\bf 95}, 261602 (2005).
\bibitem{witten}
E. Witten, J. High Energy Phys. {\bf 9807}, 006 (1998).
\bibitem{a-3q}
O. Andreev, Phys.Lett.B {\bf 756}, 6 (2016).
\bibitem{a-QQq}
O. Andreev, J. High Energy Phys. {\bf 05} (2021) 173.
\bibitem{wise}
M.J. Savage and M.B. Wise, Phys.Lett.B {\bf 248}, 177 (1990).
\bibitem{a-stb3q}
O. Andreev, Phys.Rev.D {\bf 104}, 026005 (2021).
\bibitem{a-q2}
O. Andreev, Phys.Rev.D {\bf 73}, 107901 (2006).
\bibitem{bulava}
J. Bulava, B. H\"orz, F. Knechtli, V. Koch, G. Moir, C. Morningstar, and M. Peardon, Phys.Lett.B {\bf 793} (2019) 493.
\bibitem{white}
C.D. White, Phys.Lett.B {\bf 652}, 79 (2007).
\bibitem{drum}
I.T. Drummond, Phys.Lett.B {\bf 434} (1998) 92.
\bibitem{JMR}
J.-M. Richard, A. Valcarce, and J. Vijande, Doubly-heavy baryons, tetraquarks, and related topics, a contribution to {\it "Mini-Workshop Bled 2018"}, arXiv:1811.02863 [hep-ph].
\bibitem{quigg}
E.J. Eichten and C. Quigg, Phys.Rev.Lett. {\bf 119}, 202002 (2017); M. Karliner and J.L. Rosner, Phys.Rev.Lett. {\bf 119}, 202001 (2017).
\bibitem{wagner}
P. Bicudo, K. Cichy, A. Peters, and M. Wagner, Phys.Rev.D {\bf 93}, 034501 (2016); P. Bicudo, M. Cardoso, A. Peters, M. Pflaumer, and M. Wagner, Phys.Rev.D {\bf 96}, 054510 (2017); M. Wagner, P. Bicudo, A. Peters, and S. Velten, Comparing meson-meson and diquark-antidiquark creation operators for a $\bar b\bar b ud$ tetraquark, a contribution to "The 38th International Symposium on Lattice Field Theory", LATTICE 2021, arXiv:2108.11731 [hep-lat].
\bibitem{gon}
R. Bruschini and P. Gonz\'alez, Phys.Rev.D {\bf 102}, 074002 (2020). 
\bibitem{wolfram}
See, e.g., MathWorld - A Wolfram Web Resource. https://reference.wolfram.com/language/ref/ProductLog.html.

\end{thebibliography}
\end{document}